\documentclass[paper]{JHEP}
\usepackage{cite,epsfig,graphicx}
\def\as{\alpha_{\mbox{\tiny S}}}
\def\ee{e^+e^-}
\def\qb{\bar q}
\def\tq{\tilde q}
\def\ycut{y_{\mbox{\scriptsize cut}}}
\def\yini{y_{\mbox{\scriptsize ini}}}
\def\msbar{$\overline{\rm MS}$}
\title{QCD Matrix Elements + Parton Showers}
\author{S.\ Catani$^a$\footnote{On leave of absence from INFN, Sezione di
Firenze, Florence, Italy.} , F.\ Krauss$^b$,
R.\ Kuhn$^{c,d}$ and B.R.\ Webber$^b$\\
$^a$Theory Division, CERN, 1211 Geneva 23, Switzerland\\[1mm]
$^b$Cavendish Laboratory, University of Cambridge, Cambridge CB3 0HE,
U.K.\\[1mm]
$^c$Institut f\"ur Theoretische Physik, TU Dresden, 01062 Dresden,
Germany\\[1mm]
$^d$Max Planck Institut f\"ur Physik Komplexer Systeme, 01187 Dresden,
Germany\\[1mm]
E-mail: \email{Stefano.Catani@cern.ch},
\email{krauss@hep.phy.cam.ac.uk}
\email{kuhn@theory.phy.tu-dresden.de},
\email{webber@hep.phy.cam.ac.uk}
}
\abstract{We propose a method for combining QCD matrix elements
and parton showers in Monte Carlo simulations of hadronic final states
in $\ee$ annihilation.  The matrix element and parton shower domains
are separated at some value $\yini$ of the jet resolution, defined
according to the $k_T$-clustering algorithm.  The matrix elements
are modified by Sudakov form factors and the parton showers are
subjected to a veto procedure to cancel dependence on $\yini$
to next-to-leading logarithmic accuracy.  The method provides
a leading-order description of hard multi-jet configurations
together with jet fragmentation, while avoiding the most
serious problems of double counting. We present first results of an
approximate implementation using the event generator {\tt APACIC++}.}
\preprint{CERN--TH/2000--367\\Cavendish--HEP--00/03\\hep-ph/0109231}
\begin{document}
\section{Introduction}\label{sec_intro}
The Monte Carlo simulation of multi-jet hadronic final states is
a challenging problem that has great practical importance in the
search for new physics processes at present and future colliders.
For example, the accurate simulation of 4-jet backgrounds was a central
issue in the search for the Higgs boson at LEP2, and multi-jets will be
a key ingredient in signatures of supersymmetry at the LHC.

Two extreme approaches to simulating multi-jets can be formulated
as follows. One can use the corresponding matrix elements, which are
available at leading, or in a few cases next-to-leading, order in $\as$,
with bare partons representing jets. Alternatively one can use the
parton model to generate the simplest possible final state
(e.g.\ $\ee\to q\qb$) and produce additional jets by parton showering.

In the matrix-element approach, a full simulation of the final state
is impossible unless one adds a model for the conversion of the
produced partons into hadrons. Any realistic model will include
parton showering, and then one has the problem of extra jet production
during showering and potential double counting of jet configurations.
On the other hand the pure parton shower approach gives a poor
simulation of configurations with several widely separated jets.

The interfacing of matrix-element and parton-shower event generators
is a topic of great current interest
\cite{Belyaev:2000wn,Sato:2001ae,Mangano:2001xp,Boos:2001cv}.
For earlier work on combining these approaches see
Refs.~\cite{Seymour:1995df,Andre:1998vh,Corcella:1998rs,Mrenna:1999mq,
Friberg:1999fh,Collins:2000qd,Potter:2001an,Dobbs:2001gb}.
Here we suggest a method in which the
domains of applicability of matrix elements and parton showers are
clearly separated at a given value $\yini$ of the jet resolution
variable $\ycut$, defined according to the
$k_T$-algorithm \cite{Dokshitzer:1991hj,Catani:1991hj}
for jet clustering (sometimes called the Durham algorithm).
Recall that two objects
$i$ and $j$ are resolved according to the  $k_T$-algorithm if
\begin{equation}\label{eq_yij}
y_{ij}\equiv 2\min\{E_i^2,E_j^2\}(1-\cos\theta_{ij})/Q^2>\ycut
\end{equation}
where $E_{i,j}$ are the energies of the objects, $\theta_{ij}$ is
the angle between their momenta and $Q$ is the overall
energy scale (the c.m.\ energy in $\ee$ annihilation). Two objects that are not
resolved are clustered by combining their four-momenta as $p_{(ij)}= p_i + p_j$.

The method we propose has the following features:
At $\ycut>\yini$ multi-jet cross
sections and distributions are given by matrix elements
modified by Sudakov form factors.
At $\ycut<\yini$ they are given by parton showers
subjected to a `veto' procedure, which cancels the
$\yini$ dependence of the  modified matrix elements
to next-to-leading logarithmic (NLL) accuracy.

Note that we do not attempt to give a complete description of any
configuration to next-to-leading order (NLO) in $\as$, which is
why we refer to ``combined'' rather than
``matched'' matrix elements and showers.
Procedures to combine parton showers with the matrix element corrections
due to the {\em first} (i.e. at the first relative order in $\as$)
hard multi-jet configuration were considered in
Refs.~\cite{Seymour:1995df,Andre:1998vh,Corcella:1998rs}.
Such procedures might be improved by including first-order
virtual corrections (see
Refs.~\cite{Friberg:1999fh,Collins:2000qd,Potter:2001an,Dobbs:2001gb}).
For the present, our main objective is to describe {\em any}
hard multi-jet configuration
to leading order, i.e.\ ${\cal O}(\as^{n-2})$ for
$n$ jets in $\ee$ annihilation, together with jet fragmentation
to NLL accuracy, while avoiding major problems of double
counting and/or missed phase-space regions.

In the present paper we consider the case of $\ee$ annihilation only.
In Sect.~\ref{sec_modme} we recall the NLL expressions for $\ee$ jet
rates, and show how they can be used to develop a systematic procedure
for improving the tree-level predictions of multi-parton configurations
above some jet resolution $\yini$. Then in  Sect.~\ref{sec_vetps}
we show how to combine these modified matrix-element configurations
with parton showers, in such a way that dependence on $\yini$ is
cancelled to NLL precision. In Sect.~\ref{sec_res} we show results
of an approximate Monte Carlo implementation of the above scheme,
and finally in Sect.~\ref{sec_conc} we present brief comments and
conclusions.

\section{Modified matrix elements}\label{sec_modme}
\subsection{NLL jet rates and Sudakov factors}\label{sec_nllrates}
The exclusive $\ee$ $n$-jet fractions at c.m.\ energy $Q$ and
$k_T$-resolution
\begin{equation}\label{eq_yini}
\yini=Q_1^2/Q^2
\end{equation}
are given to NLL accuracy\footnote{By NLL accuracy, we mean that the
leading and next-to-leading logarithmic contributions
$\as^n\ln^{2n}Q/Q_1$ and $\as^n\ln^{2n-1}Q/Q_1$ are included in the
expressions for $R_n(Q_1,Q)$.} for $n=2,3,4$ by~\cite{Catani:1991hj}
\begin{eqnarray}\label{eq_R2}
R_2(Q_1,Q) &=& \left[\Delta_q(Q_1,Q)\right]^2 \;, \\
\label{eq_R3}
R_3(Q_1,Q) &=& 2\left[\Delta_q(Q_1,Q)\right]^2
\int_{Q_1}^Q dq\,\Gamma_q(q,Q)\Delta_g(Q_1,q) \;, \\
\label{eq_R4}
R_4(Q_1,Q) &=& 2\left[\Delta_q(Q_1,Q)\right]^2 \Biggl\{\nonumber \\
& &\int_{Q_1}^Q dq\,\Gamma_q(q,Q)\Delta_g(Q_1,q)
   \int_{Q_1}^Q dq'\,\Gamma_q(q',Q)\Delta_g(Q_1,q')\nonumber \\
&+&\int_{Q_1}^Q dq\,\Gamma_q(q,Q)\Delta_g(Q_1,q)
   \int_{Q_1}^q dq'\,\Gamma_g(q',q)\Delta_g(Q_1,q')\nonumber \\
&+&\int_{Q_1}^Q dq\,\Gamma_q(q,Q)\Delta_g(Q_1,q)
   \int_{Q_1}^q dq'\,\Gamma_f(q')\Delta_f(Q_1,q')\Biggr\}
\end{eqnarray}
where $\Gamma_{q,g,f}$ are $q\to qg$, $g\to gg$ and $g\to q\qb$
branching probabilities
\begin{eqnarray}\label{eq_Gq}
\Gamma_q(q,Q) &=& \frac{2C_F}{\pi}\frac{\as(q)}{q}
\left(\ln\frac Q q -\frac 3 4\right) \\
\label{eq_Gg}
\Gamma_g(q,Q) &=& \frac{2C_A}{\pi}\frac{\as(q)}{q}
\left(\ln\frac Q q -\frac{11}{12}\right) \\
\label{eq_Gf}
\Gamma_f(q) &=& \frac{N_f}{3\pi}\frac{\as(q)}{q}\;,
\end{eqnarray}
$C_F=(N_c^2-1)/2N_c$ and $C_A=N_c$ for $N_c$ colours,
$N_f$ is the number of active flavours,
and $\Delta_{q,g}$ are the quark and gluon Sudakov form factors
\begin{eqnarray}\label{eq_Dq}
\Delta_q(Q_1,Q) &=& \exp\left(-\int_{Q_1}^Q dq\,\Gamma_q(q,Q)\right) \\
\label{eq_Dg}
\Delta_g(Q_1,Q) &=& \exp\left(-\int_{Q_1}^Q dq\,
\left[\Gamma_g(q,Q)+\Gamma_f(q)\right]\right)
\end{eqnarray}
with
\begin{equation}\label{eq_Df}
\Delta_f(Q_1,Q) = \left[\Delta_q(Q_1,Q)\right]^2/\Delta_g(Q_1,Q)\;.
\end{equation}
The QCD running coupling $\as(q)$ is defined in the \msbar\ renormalization
scheme. Part of the contributions beyond NLL order can be included in the
calculation by using the definition of $\as(q)$ in the bremsstrahlung scheme of
Ref.~\cite{Catani:1991rr}.

The Sudakov form factors $\Delta_i(Q_1,Q)$ for $i=q,g$
represent the probability\footnote{The NLL approximate expressions
in Eqs.~(\ref{eq_Gq},\ref{eq_Gg}) can lead to $\Delta_i>1$. In that
case one should replace $\Delta_i>1$ by 1.} for a quark or gluon to
evolve from scale $Q$ to scale $Q_1$ without any branching
(resolvable at scale $Q_1)$. Thus $R_2$ is simply the probability
that the produced quark and antiquark both evolve without branching.
More generally, the probability for a parton of type $i$ to evolve
from scale $Q$ to $q\ge Q_1$ without branching (resolvable at scale $Q_1$)
is $\Delta_i(Q_1,Q)/\Delta_i(Q_1,q)$.

\begin{figure}\begin{center}
\epsfig{file=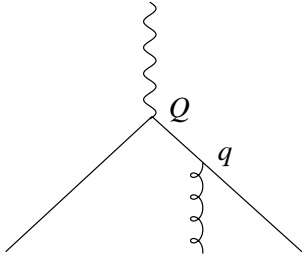,width=4cm}
\caption{Branching structure of three-jet final state.}
\label{fig_qqbarg}
\end{center}\end{figure}

In the expression (\ref{eq_R3}) for $R_3$, a gluon jet is resolved at
scale $q$ where
\begin{equation}\label{eq_q}
\min\{y_{qg},y_{\qb g}\} = q^2/Q^2\;.
\end{equation}
Recall that in coherent parton branching the evolution variable is the
emission angle \cite{cps} and the corresponding scale is the parton energy
times the angle \cite{Ellis:1996qj}. In the contribution depicted in
Fig.~\ref{fig_qqbarg}, the energy and angular regions of the phase space
that dominate at NLL order are $Q \sim E_q \sim E_{\qb} > E_g$ and
$1 \sim \theta_{q\qb} > \theta_{\qb g}$.  The quark
evolves from scale $E_q\theta_{q\qb}\sim Q$ to $Q_1$ without branching,
while the antiquark evolves from $E_{\qb}\theta_{q\qb}\sim Q$
to $\tq\sim E_{\qb}\theta_{\qb g}$ and then branches.
The resulting antiquark evolves from $\tq$ to $Q_1$, while the gluon evolves
from $q\sim E_g\theta_{\qb g}$ to $Q_1$, both without branching.
Thus the overall NLL probability is
\begin{equation}\label{eq_qqg}
\Delta_q(Q_1,Q)\frac{\Delta_q(Q_1,Q)}{\Delta_q(Q_1,\tq)}
\Gamma_q(q,Q)\Delta_q(Q_1,\tq)\Delta_g(Q_1,q)=\Gamma_q(q,Q)\,F_{q\qb g}(Q_1,Q;q)
\end{equation}
where the `Sudakov factor' $F_{q\qb g}$ is
\begin{equation}\label{eq_Fqqg}
F_{q\qb g}(Q_1,Q;q) = \left[\Delta_q(Q_1,Q)\right]^2\Delta_g(Q_1,q)\;.
\end{equation}
Taken together with the contribution in which the quark branches instead of
the antiquark, this gives Eq.~(\ref{eq_R3}) after integration over
$Q_1<q<Q$.

\begin{figure}\begin{center}\epsfig{file=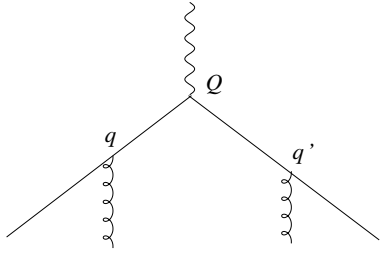,width=5cm}
\caption{An Abelian four-jet contribution.}\label{fig_clus_abel}
\end{center}\end{figure}
For four or more jets, there are several branching configurations with
different colour factors. The first term in the curly bracket of
Eq.~(\ref{eq_R4}) comes from Abelian (QED-like) contributions
such as Fig.~\ref{fig_clus_abel}, with associated probability
\begin{eqnarray}\label{eq_clus_abel}
&&\frac{\Delta_q(Q_1,Q)}{\Delta_q(Q_1,\tq)}\Gamma_q(q,Q)\Delta_q(Q_1,\tq)
\Delta_g(Q_1,q)\frac{\Delta_q(Q_1,Q)}{\Delta_q(Q_1,\tq')}
\Gamma_q(q',Q)\Delta_q(Q_1,\tq')\Delta_g(Q_1,q')\nonumber\\
&&\>=\>\Gamma_q(q,Q)\,\Gamma_q(q',Q)\,F_{q\qb gg}(Q_1,Q;q,q')
\end{eqnarray}
where the Sudakov factor is now
\begin{equation}\label{eq_Fqqgg}
F_{q\qb gg}(Q_1,Q;q,q') = \left[\Delta_q(Q_1,Q)\right]^2
\Delta_g(Q_1,q)\Delta_g(Q_1,q')\;.
\end{equation}

\begin{figure}\begin{center}\epsfig{file=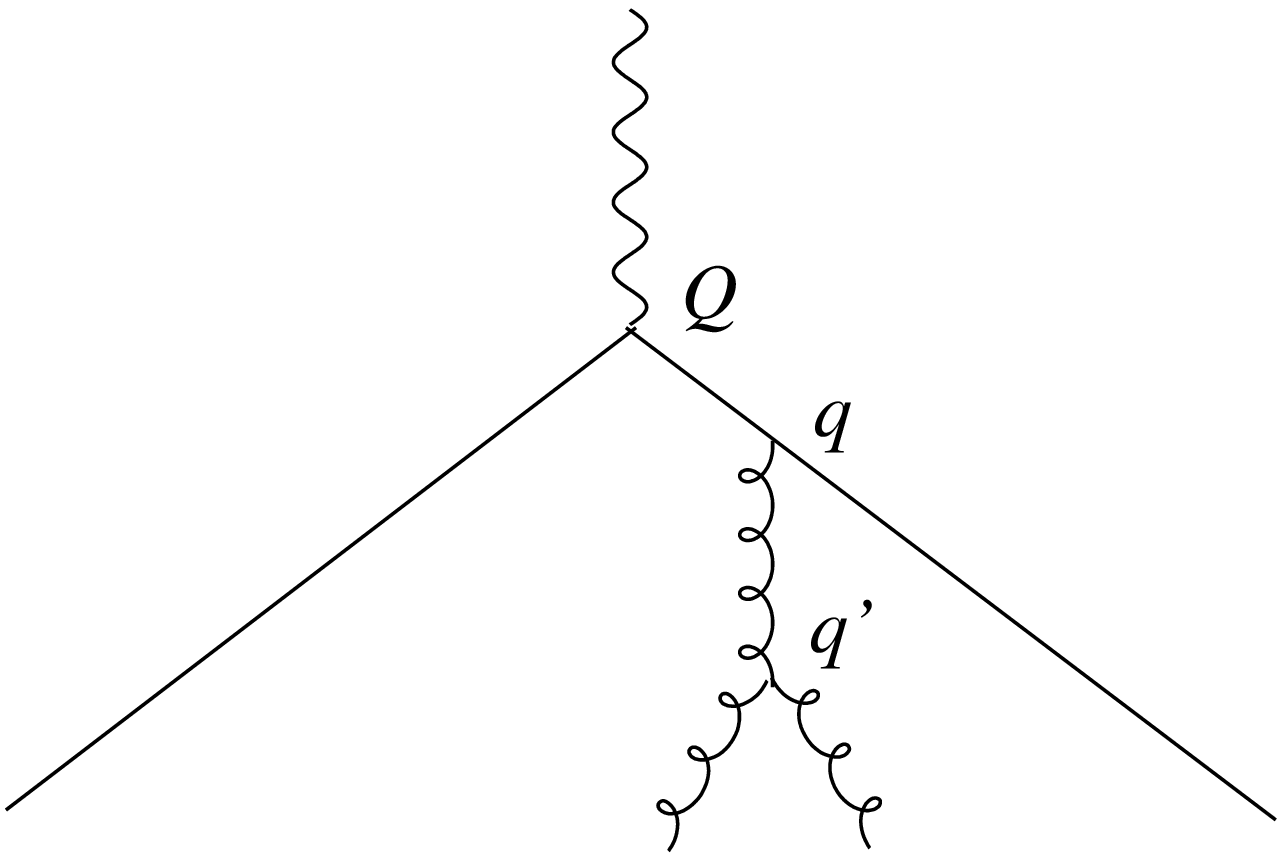,width=5cm}
\caption{A non-Abelian four-jet contribution.}\label{fig_tree}
\end{center}\end{figure}

The second term in the curly bracket of Eq.~(\ref{eq_R4}) comes from
contributions with a $q\to q g$ branching at scale $q$ followed by
$g\to g g$ at scale $q'$ (Fig.~\ref{fig_tree}).
The probability of this is
\begin{eqnarray}\label{eq_clus_nab}
&&\Delta_q(Q_1,Q)\frac{\Delta_q(Q_1,Q)}{\Delta_q(Q_1,\tq)}\Gamma_q(q,Q)
\Delta_q(Q_1,\tq)\frac{\Delta_g(Q_1,q)}{\Delta_g(Q_1,\tq')}
\Gamma_g(q',q)\Delta_g(Q_1,q')\Delta_g(Q_1,\tq')\nonumber\\
&&\>=\>\Gamma_q(q,Q)\,\Gamma_g(q',q)\,F_{q\qb gg}(Q_1,Q;q,q')
\end{eqnarray}
where the factor $F_{q\qb gg}$ is {\em the same} as that given in
Eq.~(\ref{eq_Fqqgg}).

The final term in Eq.~(\ref{eq_R4}) corresponds to diagrams like
Fig.~\ref{fig_tree} except that the branching at $q'$ is $g\to q\qb$
instead of $g\to gg$.   The factor of $\Gamma_g(q',q)$ is replaced by
$\Gamma_f(q')$ given by Eq.~(\ref{eq_Gf}), and $\Delta_g(Q_1,q')$
becomes $\Delta_f(Q_1,q')$ given by Eq.~(\ref{eq_Df}).  Thus the
Sudakov factor becomes
\begin{equation}\label{eq_Fqqgf}
F_{q\qb q\qb}(Q_1,Q;q,q') = \left[\Delta_q(Q_1,Q)\right]^2
\Delta_g(Q_1,q)\Delta_f(Q_1,q')\;.
\end{equation}

We see that in general the overall Sudakov factor depends on the
{\em nodal values} of the $k_T$-scale $q,q',\ldots$ at which branching
occurs, and on the types of partons involved. There is an overall
factor of $\left[\Delta_q(Q_1,Q)\right]^2$ coming from $q\qb$
production at scale $Q$, a factor of
$\Delta_g(Q_1,q)$ when a gluon is emitted at scale $q$, and a factor
$\Delta_f(Q_1,q)$ when a gluon branches to quark-antiquark at scale $q$.
Although we have explicitly discussed only the $n=2,3,4$ jet rates,
this structure of the Sudakov factor is valid for any $n$, as can be
derived from the generating function given in Ref.~\cite{Catani:1991hj}.

\subsection{Matrix element improvement}
We can improve the description of the 3-jet distribution throughout the
region $y_{q\qb}>y_{qg},y_{\qb g}>\yini$ by using the full tree-level matrix
element squared  $|{\cal M}_{q\qb g}|^2$ in place of the NLL branching
probability $\Gamma_q(q,Q)$ in Eq.~(\ref{eq_qqg}).
More precisely, we generate $q\qb g$ momentum configurations
according to the matrix element squared, with resolution cutoff
$\yini=Q_1^2/Q^2$, and then weight each configuration by the
Sudakov factor $F_{q\qb g}(Q_1,Q;q)$ in Eq.~(\ref{eq_Fqqg}),
where $q$ is given by Eq.~(\ref{eq_q}). For consistency with
Eqs.~(\ref{eq_Gq})--(\ref{eq_Gf}), we should also use $q$ as the
argument of the running coupling in the matrix element squared.

Similarly in the four-jet case of Eq.~(\ref{eq_clus_abel})
the product $\Gamma_q(q,Q)\Gamma_q(q',Q)$ is an approximation
to the full matrix element squared $|{\cal M}_{q\qb gg}|^2$
in  the kinematic region where $y_{qg}$ and $y_{\qb g'}$
are the smallest interparton separations. Thus it is legitimate
in NLL approximation to replace it by $|{\cal M}_{q\qb gg}|^2$
in that region.  The remaining factor $F_{q\qb gg}(Q_1,Q;q,q')$ in
Eq.~(\ref{eq_clus_abel}) is the extra Sudakov weight to be applied.

In general, we obtain an improved description of the jet rates and
distributions, above the resolution value $\yini$, by choosing the parton
configurations according to the tree-level matrix elements squared
and then weighting them by a product of Sudakov form factors. The
arguments of the form factors and the running coupling
are given by the nodal values of the
$k_T$-resolution in the branching process, estimated by applying
the $k_T$-clustering algorithm to the parton configuration.

\subsection{General procedure}\label{sec_proc}
The proposed procedure for generating $\ee\to n$-jet configurations at
c.m.\ energy $Q$ and jet resolution $\yini$ is thus as follows:
\begin{enumerate}
\item
Select the jet multiplicity $n$ and parton identities $i$ with probability
\begin{equation}\label{eq_Pn}
P^{(0)}(n,i) = \frac{\sigma^{(0)}_{n,i}}{\sum_{k,j}^{k=N}\sigma^{(0)}_{k,j}}
\end{equation}
where $\sigma^{(0)}_{n,i}$ is the tree-level $\ee\to n$-jet cross section
at resolution $\yini=Q_1^2/Q^2$, calculated using a fixed value
$\as(Q_1)$ for the strong coupling. The label $i$ is to distinguish
different parton identities with the same multiplicity,
e.g.\ $i=q\qb gg$ or $q\qb q\qb$ for $n=4$.
$N$ is the largest jet multiplicity for which the
calculation can realistically be performed ($N\sim 6$ currently). Errors
will then be of relative order $\as^{N-1}$. Ideally,
one should check that any given result is insensitive to $N$.
\item
Distribute the jet momenta according to the corresponding $n$-parton
matrix elements squared $|{\cal M}_{n,i}|^2$, again using fixed $\as(Q_1)$.
\item
Use the $k_T$-clustering algorithm to determine the
resolution values $y_2=1>y_3>\ldots>y_n>\yini$ at which $2,3,\ldots,n$
jets are resolved. These give the nodal values of $q_j=Q\sqrt{y_j}$
for a tree diagram that specifies the $k_T$-clustering sequence for that
configuration.
\item Apply a coupling-constant weight of
$\as(q_3)\as(q_4)\cdots\as(q_n)/[\as(Q_1)]^{n-2}<1$.
\item
For each internal line of type $i$ from a node at scale $q_j$ to the
next node at $q_k<q_j$, apply a Sudakov weight factor
$\Delta_i(Q_1,q_j)/\Delta_i(Q_1,q_k)<1$. For an external line from a node
at scale $q_j$, the weight factor is $\Delta_i(Q_1,q_j)$.
This procedure gives the overall Sudakov factors
$F_i(Q_1,Q;q_3,\ldots,q_n)$ of Sect.~\ref{sec_nllrates}.
\item
Accept the configuration if the product of the coupling-constant weight
and the Sudakov factor is greater than a random number ${\cal R}\in [0,1]$
times\footnote{Multiplying by $[\Delta_q(Q_1,Q)]^2$ increases the efficiency
of the procedure, since this constant factor is always present.}
$[\Delta_q(Q_1,Q)]^2$.  Otherwise, return to step 1.
\end{enumerate}
Note that the weight assignment is a fully gauge-invariant
procedure relying only on the types (quark or gluon) and momenta of the
final-state partons. The weight factor is actually independent of
the detailed structure of the clustering tree and is the same as
that for the Abelian (QED-like) graph with the same nodal scale
values: see, for example, Eqs.~(\ref{eq_clus_abel}) and
(\ref{eq_clus_nab}).

An advantage of the above procedure is that it adjusts the jet
multiplicity distribution to include the Sudakov and coupling-constant
weights, without the need for separate numerical integrations.  To
prove this, note that the probability of accepting an $(n,i)$-parton final
state, once selected, is $p_{n,i}=\sigma_{n,i}/\sigma^{(0)}_{n,i}$,
where $\sigma_{n,i}$ includes the weight factors. The overall probability
$P(n,i)$ of selecting an $(n,i)$-parton state is the probability of rejecting
any state any number of times before finally accepting the $(n,i)$ state. Thus
\begin{eqnarray}
P(n,i) &=& \sum_{m=0}^\infty\left[\sum_{k,j}^{k=N}P^{(0)}_{k,j}
(1-p_{k,j})\right]^m P^{(0)}_{n,i}p_{n,i}\nonumber \\
&=& \frac{P^{(0)}_{n,i}p_{n,i}}{\sum_{k,j}^{k=N}P^{(0)}_{k,j}p_{k,j}}
\>=\>\frac{\sigma_{n,i}}{\sum_{k,j}^{k=N}\sigma_{k,j}}\;,
\end{eqnarray}
as required.

In the clustering step 3, attempted clustering of partons will
sometimes be `wrong': for example,
a $q\qb g$ final state may be clustered first as $(q\qb)g$. The nodal value
for the $(q\qb)$ clustering is irrelevant to NLL accuracy since there is no
associated soft or collinear enhancement. Hence the optimal procedure
is to forbid such a clustering and continue until either $(qg)$ or
$(\qb g)$ is clustered.  In more complicated cases, e.g.\ $q\qb q\qb$,
the clustering $(q\qb)$ is allowed but $(qq)$ and $(\qb\qb)$ should always
be forbidden. This is simply achieved by moving to the pair of objects
with the next-higher value of $y_{ij}$ whenever the lowest value
belongs to a forbidden combination.

\section{Vetoed Parton Showers}\label{sec_vetps}
\subsection{Angular ordering and veto procedure}
Having generated multi-jet distributions above the resolution value
$\yini$ according to matrix elements modified by form factors,
it remains to generate distributions at lower values of $\ycut$
by means of parton showers. This should be done in such a way that
the dominant (LL and NLL) dependence on the arbitrary parameter $\yini$
cancels. Any residual dependence on $\yini$ could be exploited for
tuning less singular terms to obtain optimal agreement with data.

Note that  $\yini$ must set an upper limit on interparton
separations $y_{ij}$ generated in the showers. Otherwise the
exclusive jet rates at resolution $\yini$ could be changed
by showering. At first sight, this might suggest that we should
evolve the showers from the scale $Q_1=Q\sqrt{\yini}$
instead of $Q$. However, this would correspond to using transverse
momentum rather than angle as the evolution variable, and therefore
it would not lead to cancellation of the dependence on $\ln\yini$.

Consider, for example, the 2-jet rate at resolution $y_0=Q_0^2/Q^2<\yini$.
If we start from $R_2$ at scale $Q_1$ and then evolve from $Q_1$ to $Q_0$,
we obtain a 2-jet rate of
\begin{equation}
\left[\Delta_q(Q_1,Q)\Delta_q(Q_0,Q_1)\right]^2
\end{equation}
instead of the correct result
\begin{equation}\label{eq_R20}
R_2(Q_0,Q) = \left[\Delta_q(Q_0,Q)\right]^2\;.
\end{equation}
This is because, although the $y_{ij}$ values in the showers
are limited by  $\yini$, the angular regions in which they evolve should
still correspond to scale (energy times angle) $Q$ rather than $Q_1$.
Consequently we should allow the showers to evolve from scale $Q$
but {\em veto} any branching with transverse momentum $q>Q_1$,
i.e.\ the selected parton branching is forbidden but that parton
has its scale reset to the current value as an upper limit for
subsequent branching.

\begin{figure}\begin{center}\epsfig{file=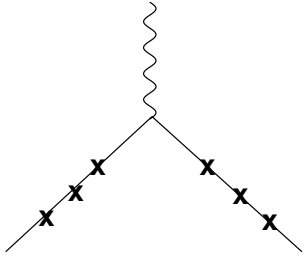,width=4cm}
\caption{Vetoed showers on two-jet contribution.}\label{fig_veto1}
\end{center}\end{figure}
The 2-jet rate at any scale $Q_0<Q_1$ is now
given by the sum of probabilities of $0,1,2,\ldots$ vetoed branchings
(represented by crosses in Fig.~\ref{fig_veto1}) and no actual resolved
branchings. The sum of these probabilities for the quark line is
\begin{eqnarray}
&&\Delta_q(Q_1,Q)\Delta_q(Q_0,Q)\left\{1+\int_{Q_1}^Q dq\,\Gamma_q(q,Q)
+ \int_{Q_1}^Q dq\,\Gamma_q(q,Q)\int_{Q_1}^q dq'\,\Gamma_q(q',Q)
+\cdots\right\}\nonumber\\
&&\>=\>\Delta_q(Q_1,Q)\Delta_q(Q_0,Q)
\exp\left(\int_{Q_1}^Q dq\,\Gamma_q(q,Q)\right)\;.
\end{eqnarray}
Comparing with Eq.~(\ref{eq_Dq}), we see that the
series sums to $1/\Delta_q(Q_1,Q)$,  cancelling the $\yini$
dependence and giving $\Delta_q(Q_0,Q)$. Similarly for the antiquark line,
so that the product does indeed give Eq.~(\ref{eq_R20}).

\begin{figure}\begin{center}\epsfig{file=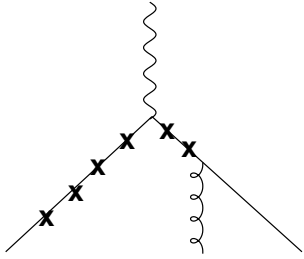,width=4cm}
\caption{Vetoed showers on contribution with two jets at scale $Q_1$
and three at scale $Q_0$.}\label{fig_veto2}
\end{center}\end{figure}

\begin{figure}\begin{center}\epsfig{file=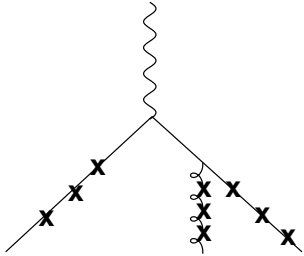,width=4cm}
\caption{Vetoed showers on contribution with three jets at scales
$Q_1$ and $Q_0$.}\label{fig_veto3}
\end{center}\end{figure}

For the 3-jet rate at scale $Q_0<Q_1$ there are two possibilities:
either the event is a 2-jet at scale $Q_1$ and then has one branching
resolved at scale $Q_0$, or it is a 3-jet at scale $Q_1$ and remains
so at scale $Q_0$. The first case is depicted in Fig.~\ref{fig_veto2}.
Its probability is
\begin{equation}
2[\Delta_q(Q_1,Q)]^2\left[\frac{\Delta_q(Q_0,Q)}{\Delta_q(Q_1,Q)}\right]^2
\int_{Q_0}^{Q_1} dq\,\Gamma_q(q,Q)\Delta_g(Q_0,q)
\end{equation}
while that of the second case (Fig.~\ref{fig_veto3}) is
\begin{equation}\label{eq_veto3}
2[\Delta_q(Q_1,Q)]^2\left[\frac{\Delta_q(Q_0,Q)}{\Delta_q(Q_1,Q)}\right]^2
\int_{Q_1}^Q dq\,\Gamma_q(q,Q)\Delta_g(Q_1,q)
\frac{\Delta_g(Q_0,q)}{\Delta_g(Q_1,q)}\;.
\end{equation}
The sum is indeed $\yini$-independent and equal to $R_3(Q_0,Q)$ as
given in Eq.~(\ref{eq_R3}). Similarly for higher jet multiplicities.
A general proof of the cancellation of  $\yini$-dependence to NLL
accuracy is given in Sect.~\ref{sec_proof}.

\subsection{Initial conditions for showers}\label{sec_ini}
Notice in Eq.~(\ref{eq_veto3}) that the vetoed parton shower from a
gluon created in a branching at scale $q>Q_1$ starts at scale $q$ rather
than $Q$ or $Q_1$. On the other hand, the shower from the quark
line starts at scale $Q$.  In general, each vetoed shower on an external
parton line must start at the scale value of the node at which that
parton was `created', in order to cancel the $Q_1$ dependence of the
associated Sudakov factor. In the case of the branching $g\to gg$,
the softer of the two gluons should be regarded as the one `created',
the harder one being traced back to a node at a higher scale.

The correct treatment of the branching $g\to q\qb$ is more subtle, although
less crucial because this branching contributes only at NLL
level.  The associated factor $\Delta_f(Q_1,q')$ in
Eq.~(\ref{eq_R4}) is a correction rather than a form factor, representing
the conversion of a gluon jet into two quark jets at scale
$q'$. Consequently the optimal treatment would be as follows: for a
$q\qb$ pair clustered at scale $q'$, coming from an internal gluon line
`created' at scale $q>q'$, one should generate a vetoed shower from the
gluon starting from scale $q$ and evolving the harder gluon at each
branching\footnote{The softer gluon, on the other hand, is allowed to
evolve down to the shower cut-off $Q_0$.} down to scale $q'$, then switch
to separate showers from the quark and antiquark starting at scale $q'$.
If this seems unnecessarily complicated for a next-to-leading contribution,
one may instead consider treating the quark and antiquark as being
`created' at the higher scale $q$ of their parent gluon. Then the colour
factor which should be $C_A$ between scales $q$ and $q'$ is approximated
by $2C_F$, an error of relative order $1/N_c^2$ in a contribution that
is already non-leading with respect to $\ln\yini$.

\subsection{Proof of cancellation of $\yini$ dependence}\label{sec_proof}
Here we make use of the generating function formalism and results
of Ref.~\cite{Catani:1991hj} to prove the cancellation of
$\yini$-dependence at NLL order.  Recall that the NLL jet fractions at
$k_T$-resolution $\yini=Q_1^2/Q^2$ in a quark jet initiated at scale $Q$
are given by
\begin{equation}\label{eq_Rqn}
R^{(q)}_n(\yini=Q_1^2/Q^2) = \frac{1}{n!}
\left(\frac{\partial}{\partial u}\right)^n
\left.\phi_q(Q_1,Q;u_q=u_g=u)\right|_{u=0}
\end{equation}
where the quark-jet generating function $\phi_q$ is \cite{Catani:1991hj}
\begin{equation}\label{eq_phiq_Q1}
\phi_q(Q_1,Q;u_q,u_g) = u_q\exp\left\{\int_{Q_1}^Q dq\,\Gamma_q(q,Q)
\left[\phi_g(Q_1,q;u_q,u_g)-1\right]\right\}\;,
\end{equation}
$\phi_g$ being the corresponding gluon-jet generating function.
Now we wish to generate the jet fractions at some lower resolution
value $\ycut=Q_0^2/Q^2 < \yini$. This is to be done by replacing $u_i$
everywhere in Eq.~(\ref{eq_phiq_Q1}) by a modified generating function
$\tilde\phi_i(Q_0,Q_1,Q;u_q,u_g)$, representing the vetoed parton
shower.  To have the correct jet fractions at scale $Q_0$ we require that
\begin{equation}\label{eq_tildephi}
\phi_i(Q_1,Q;\tilde\phi_q,\tilde\phi_g)=\phi_i(Q_0,Q;u_q,u_g)\;.
\end{equation}
Consequently we must have
\begin{equation}\label{eq_phiq_Q0}
\phi_q(Q_0,Q;u_q,u_g) = \tilde\phi_q(Q_0,Q_1,Q;u_q,u_g)
\exp\left\{\int_{Q_1}^Q dq\,\Gamma_q(q,Q)
\left[\phi_g(Q_0,q;u_q,u_g)-1\right]\right\}\;.
\end{equation}
Hence
\begin{eqnarray}\label{eq_phiq_veto}
\tilde\phi_q(Q_0,Q_1,Q;u_q,u_g)&=&\phi_q(Q_0,Q;u_q,u_g)
\exp\left\{-\int_{Q_1}^Q dq\,\Gamma_q(q,Q)
\left[\phi_g(Q_0,q;u_q,u_g)-1\right]\right\}\nonumber\\
&=&u_q\exp\left\{\int_{Q_0}^{Q_1} dq\,\Gamma_q(q,Q)
\left[\phi_g(Q_0,q;u_q,u_g)-1\right]\right\}\;,
\end{eqnarray}
using Eq.~(\ref{eq_phiq_Q1}) with $Q_1$ replaced by $Q_0$
for $\phi_q(Q_0,Q;u_q,u_g)$.  Thus the modified
generating function $\tilde\phi_q(Q_0,Q_1,Q;u_q,u_g)$ differs from the full
generating function $\phi_q(Q_0,Q;u_q,u_g)$ only by having $Q_1$ as the
upper limit on the $q$-integration in place of $Q$, i.e.\ by having a
veto, $q<Q_1$. Note that $Q$ remains $Q$ in the integrand $\Gamma_q$,
so this is {\em not} equivalent to an unvetoed secondary shower starting
at scale $Q_1$. Note also that $Q$ is the initial scale of the quark-jet
generating function in Eq.~(\ref{eq_phiq_Q1}): 
as pointed out in Sect.~\ref{sec_ini}, 
this is the scale value of the node at which the external quark is `created'.

A similar result holds for gluon jets. The only difference between quark and
gluon jets concerns the treatment of the branching $g \to q\qb$, as discussed in
Sect.~\ref{sec_ini}.

\subsection{Colour structure}
\begin{figure}\begin{center}\epsfig{file=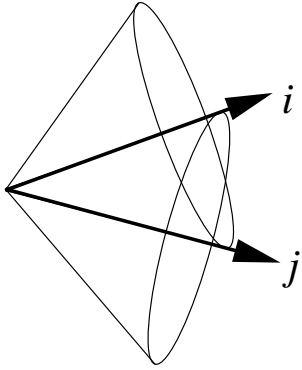,width=4cm}
\caption{Parton shower cones.}\label{fig_cones}
\end{center}\end{figure}

The vetoed shower from each parton evolves in the
{\em phase space for angular-ordered branching} \cite{Marchesini:1988cf}.
This depends on the colour structure of
the matrix element.  As illustrated in Fig.~\ref{fig_cones},
the angular region for parton $i$ is a cone bounded by the direction of
parton $j$ (and vice-versa), where $i$ and $j$ are colour-connected.
The upper limit on the scale in the vetoed shower for each parton is
given by the energy of that parton times the relevant cone angle. 
This prescription identifies the cone angles for the `intrinsic' radiation
from each parton and is correct when the matrix element describes parton
configurations at a hard scale $Q_1 \sim Q$.

However, in our case some of the hard partons are produced at the scale $Q$,
which is much larger than the resolution scale $Q_1$, and the relevant cone
angles are not set directly by the final state at scale $Q_1$.
To cancel the dependence on the logarithms of $Q/Q_1$ to NLL precision, the
vetoed shower has to include `interparton' radiation \cite{intrad},
i.e.\ soft gluons emitted at angles that are larger than the
cone angles for `intrinsic' radiation. 
In the $q\qb g$ case depicted in Fig.~\ref{fig_qqbarg}, for example, the
nodal scale is $q\sim E_g\theta_{\qb g}$.
The vetoed shower from the antiquark has to include not only gluons emitted
at smaller angles $\theta_g<\theta_{\qb g}$ but also those emitted at larger
angles, $\theta_{\qb g}<\theta_g<\theta_{q\qb}$, with energies
less than $Q_1/\theta_g$. These soft gluons emitted at large angles are
radiated {\em coherently} by the final-state gluon and antiquark.
Thus the cone angle for the vetoed antiquark shower
is $\theta_{q\qb}$ and the initial scale is $E_{\qb}\theta_{q\qb}\sim Q$.

Notice that the starting conditions for the vetoed showers are
deduced from the application of the $k_T$-clustering algorithm to the parton
configurations generated from the modified matrix elements.
It is not necessary to assign a colour structure explicitly to the
final state at scale $Q_1$ for this purpose.  The relevant
colour structures are sampled with the correct probabilities to
cancel $\yini$-dependence to NLL order. On the other hand, if a
hadronization model (cluster or string) is to be applied after the
showers, a specific colour connection structure must be provided
to the model.

If the colour structure is not unique, colour connections can be
selected according to their relative contributions to the matrix
element squared, which are well-defined in the limit that the number
of colours $N_c$ is large. Corrections to the large-$N_c$ limit are normally
of relative order $1/N_c^2$.
For high parton multiplicity, when the colour structure is not easily
computable even at large $N_c$, one may use the clustering scheme
as a first approximation in assigning colour connections. This is the
procedure we shall adopt in Sect.~\ref{sec_res}.

\section{Results}\label{sec_res}
An approximate version of the procedure described above has been implemented
in version 1.1 of the event generator {\tt APACIC++} \cite{Krauss:1999fc}
as follows:
\begin{enumerate}
\item{Cross sections $\sigma^{(0)}_n$ for the production of 2, 3, 4,
      and 5 jets according to
      some $\yini$ are calculated at the tree--level. The tree--level cross
      sections are translated into rates via
      \begin{equation}\label{eq_R345}
      {\cal R}_{3,4,5}(\yini) =
           \frac{\sigma^{(0)}_{3,4,5}(\yini)}{\sigma^{(0)}_2}\;,\;\;\;
      {\cal R}_2(\yini) = 1-\sum\limits_{i=3}^5 {\cal R}_i(\yini)\;.
      \end{equation}
      For each number 3, 4, and 5 of jets, the argument of $\alpha_s$ is
      chosen to be $\kappa_i Q^2$, where the factors $\kappa_{3,4,5}$ are
      adjustable parameters chosen to reproduce the measured jet
      rates. Note that this determination of the jet rates is slightly
      different from the one outlined in Sect.~\ref{sec_proc}, for simplicity
      and to allow extra freedom in fitting the measured rates.}
\item{The number of partons and their flavours are now chosen according
      to the corresponding rates in Eq.~(\ref{eq_R345}).}
\item{The four--momenta of the jets are generated
      according to the appropriate tree--level matrix element.}
\item{The $k_T$--clustering algorithm is applied sequentially
      until only two jets remain. The event is accepted with
      probability equal to the weight assigned to the
      sequence of clustering, computed as described in
      points 4 and 5 of Sect.~\ref{sec_proc}.
      As recommended there, the remaining two jets are `forced'
      to be a quark--antiquark pair.
      When an event is rejected, a new configuration of
      momenta is chosen, i.e.\ the program returns to step 3.}
\item{Next the colour configuration is chosen to be identical to the
      topology obtained in the clustering step above.}
\item Finally, parton showers are generated on external lines according to
the {\tt APACIC++} algorithm described in Ref.~\cite{Krauss:1999fc},
except that a veto on emission with transverse momentum greater than $Q_1$
is applied. In {\tt APACIC++}, the evolution variable is virtuality and
angular ordering is imposed. The initial conditions on the showers appear
somewhat more restrictive than those proposed in Sect.~\ref{sec_ini}, and
so a slight reduction in QCD radiation is expected in this approximate
implementation of the veto procedure.
\end{enumerate}

Note that within {\tt APACIC++}, more options for the steps outlined
above exist, which are described in some detail in the
manual \cite{Krauss:1999fc}.
For instance, jet rates can be chosen according to the NLL--rates
of Eqs.~(\ref{eq_R2}--\ref{eq_R4}), in clustering to two jets the
configuration can be rejected if the two remaining jet flavours do not
correspond to an quark--antiquark pair, and
the colour configuration of the jets can be chosen in a probabilistic
fashion following the prescription of Ref.~\cite{Andre:1998vh}.

However, we find at present that the procedure above yields the best
agreement with experimental data. It leaves a number of parameters to
be tuned, namely
\begin{enumerate}
\item{The value of $\alpha_s$ at some reference scale. We have chosen the
      scale of LEP 1, the $Z$--pole. For the results displayed in
      the Figures, $\as=0.1127$ was found in the tune of Ref.~\cite{Uwe}.}
\item{The value of the jet resolution parameter $\yini$ at which one
      divides the phase space into a region populated by the matrix
      elements and the region populated by the parton showers. The weak
      (beyond NLL) dependence on this parameter
      has been employed for optimizing agreement with
      data. In the tune, the value of $\yini$ was fixed to
      $\yini = 10^{-2.4}$.}
\item{The values of the three scale factors $\kappa_{3,4,5}$. These are
      supposed to compensate to some extent for the absence
      of subleading corrections to jet rates at the parton level.
      The tune gave $\kappa_{3,4,5} = 10^{-1.35,-1.48,-3.08}$.}
\end{enumerate}

The parameters above together with the infrared cut--off of the parton
shower and some fragmentation parameters have been tuned recently; for more
details we refer to \cite{Uwe}. In the following we display some
illustrative results, comparing the performance of {\tt APACIC++}
with the standard event generators {\tt HERWIG} \cite{Marchesini:1992ch},
{\tt PYTHIA} \cite{Sjostrand:2000wi},
{\tt ARIADNE} \cite{Lonnblad:1992tz} and
with data taken by the DELPHI collaboration.
The parameters of {\tt HERWIG}, {\tt PYTHIA} and {\tt ARIADNE} were
tuned in Refs.~\cite{Tune3}, \cite{Tune2} and \cite{Tune1}, respectively.

\begin{figure}
\begin{center}
\begin{tabular}{cc}
\includegraphics[width=6.5cm,height=6.5cm]{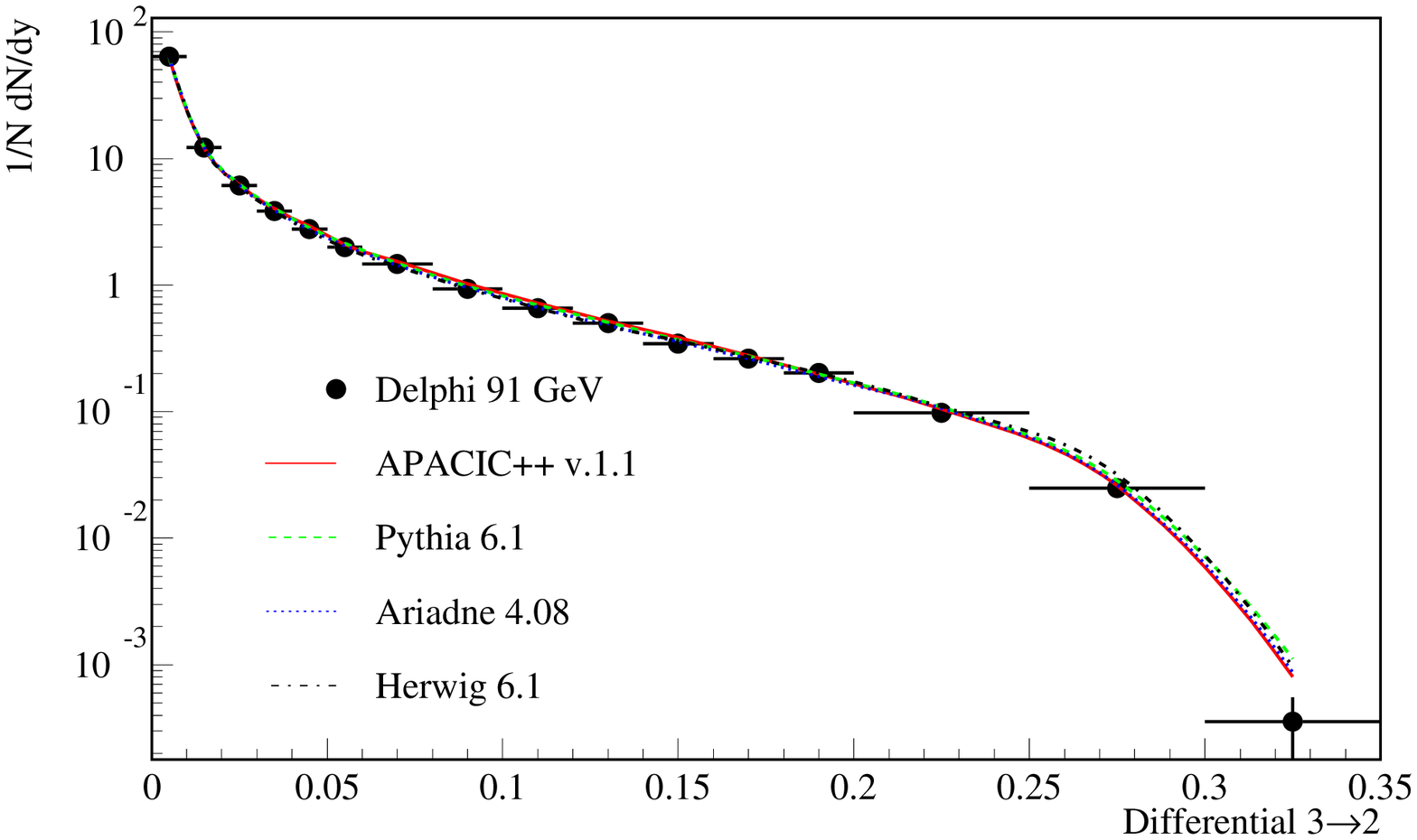} &
\includegraphics[width=6.5cm,height=6.5cm]{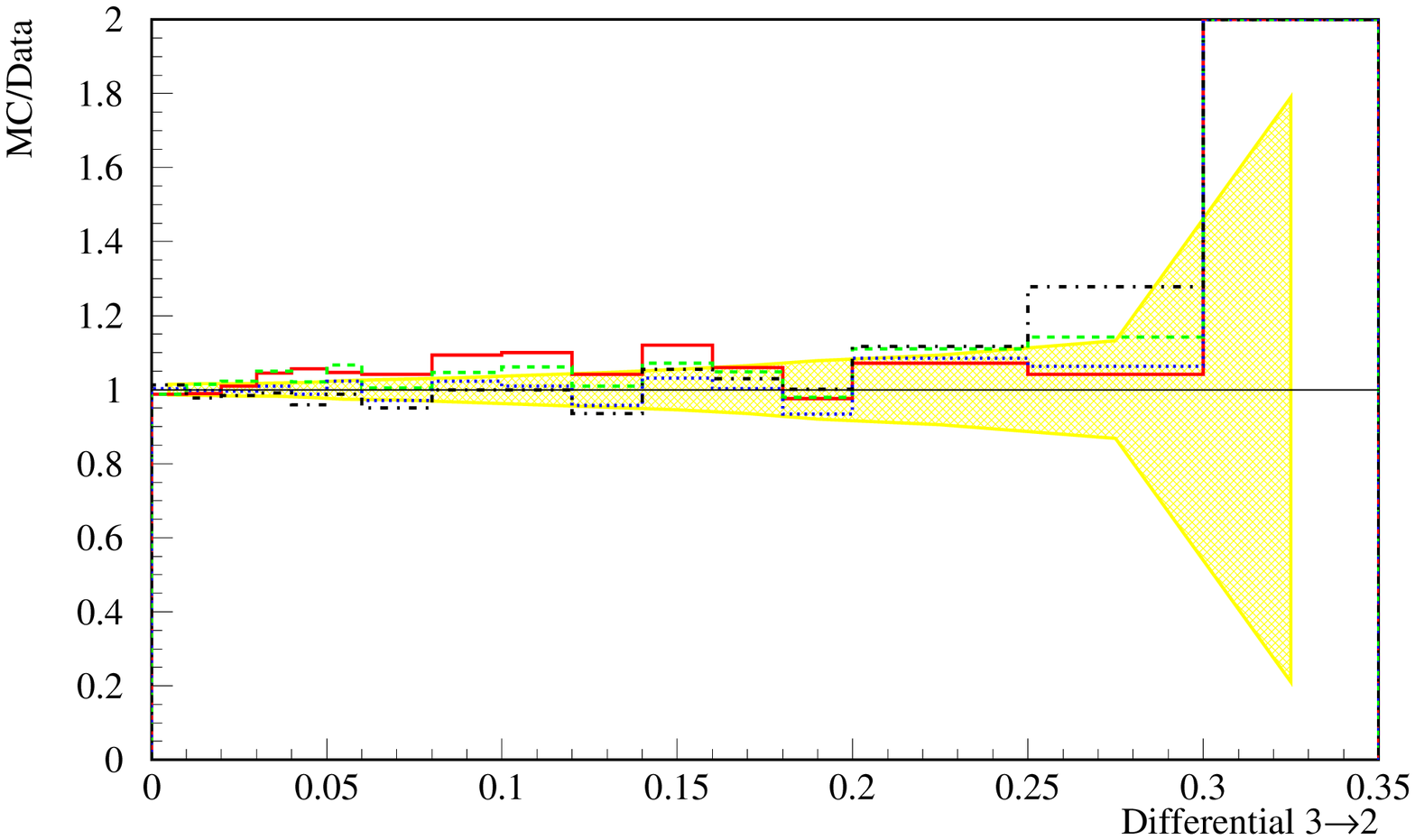} \\
\includegraphics[width=6.5cm,height=6.5cm]{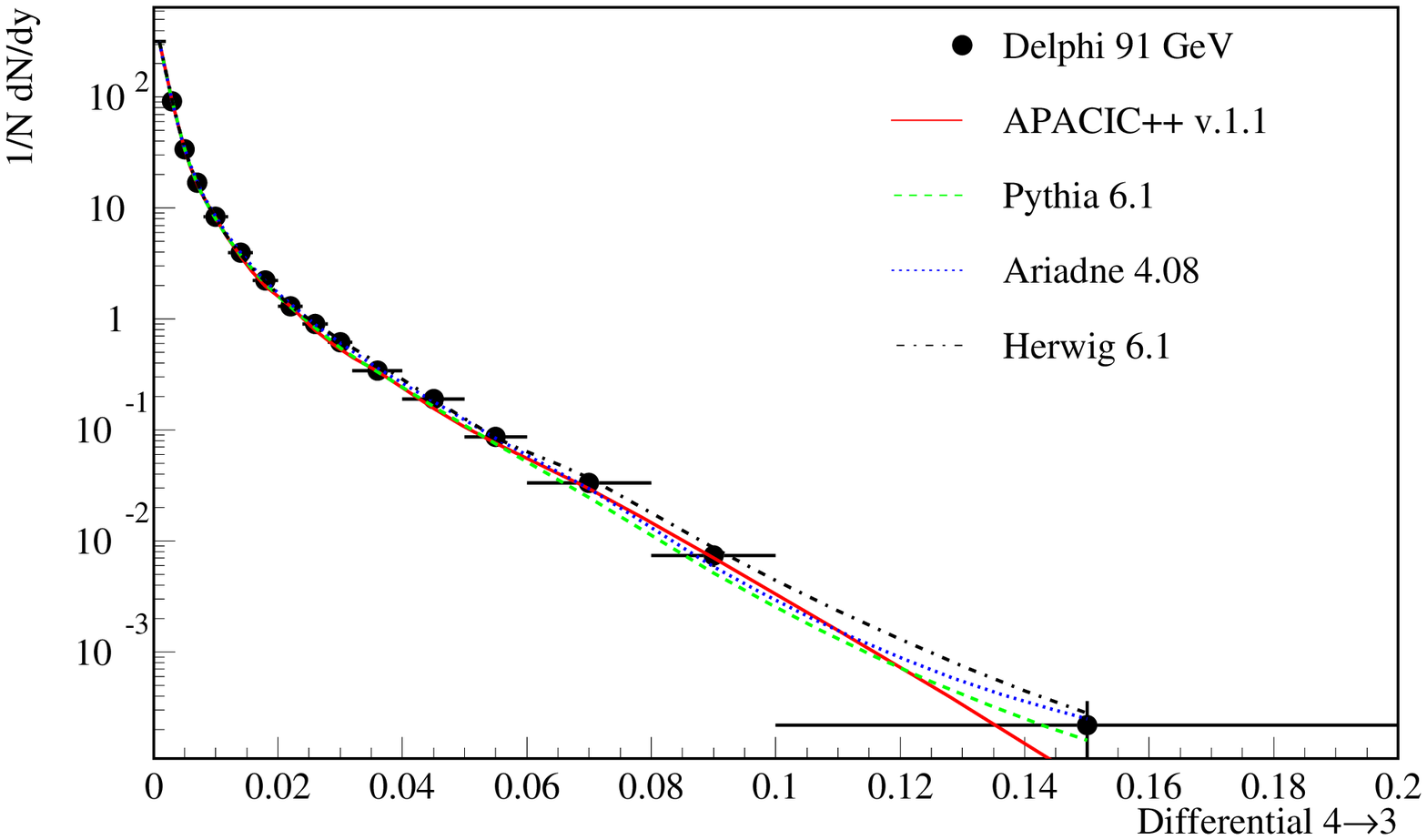} &
\includegraphics[width=6.5cm,height=6.5cm]{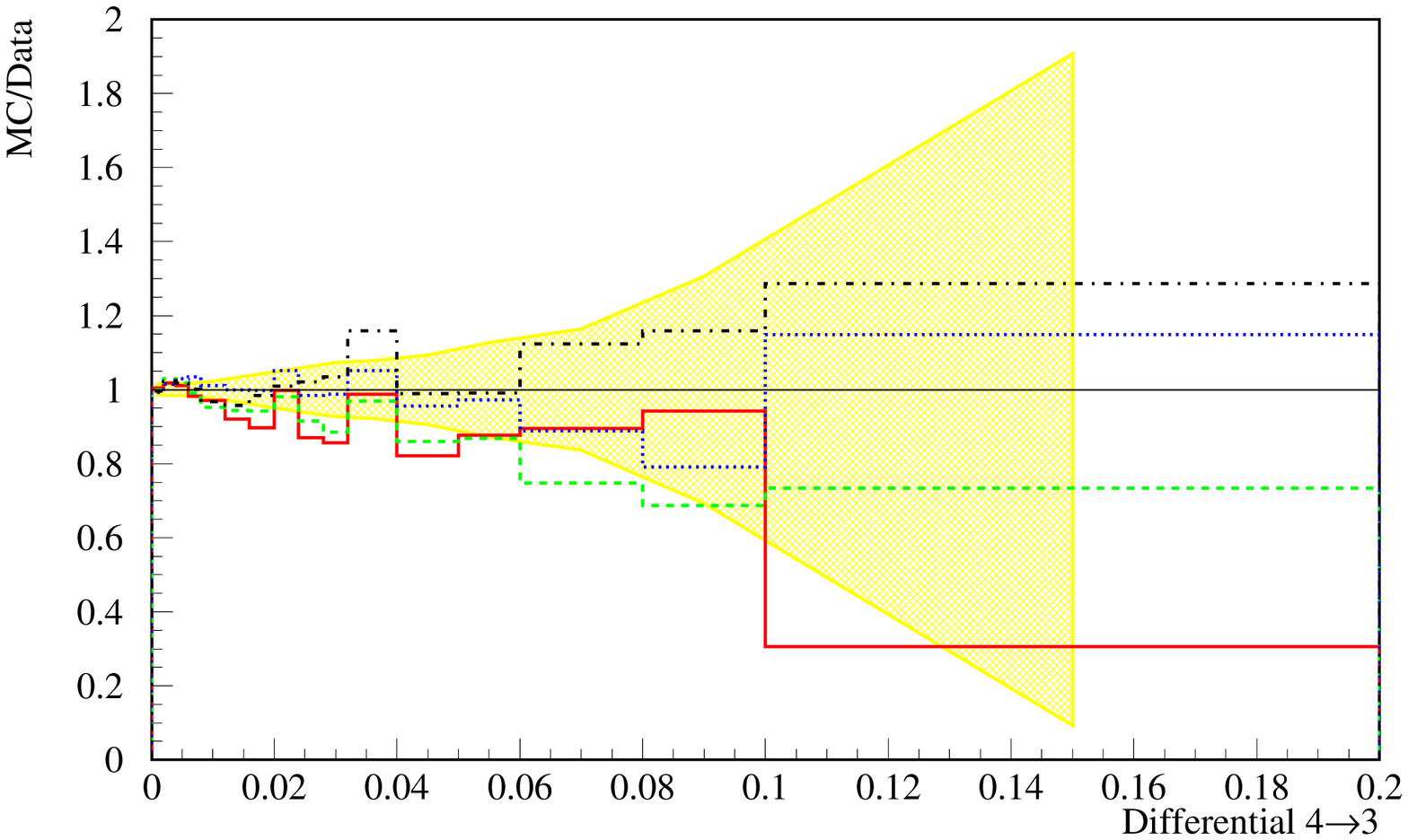} \\
\includegraphics[width=6.5cm,height=6.5cm]{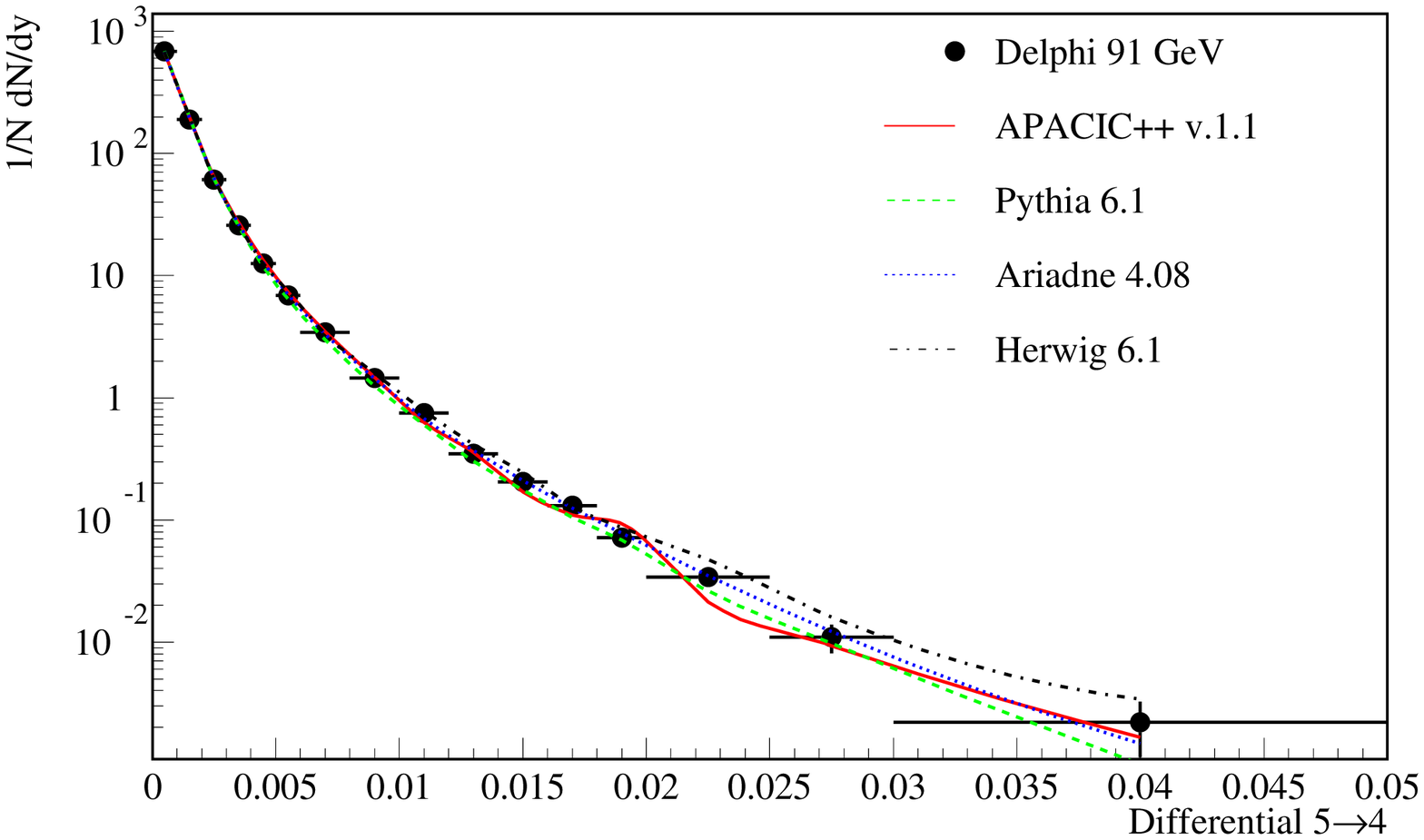} &
\includegraphics[width=6.5cm,height=6.5cm]{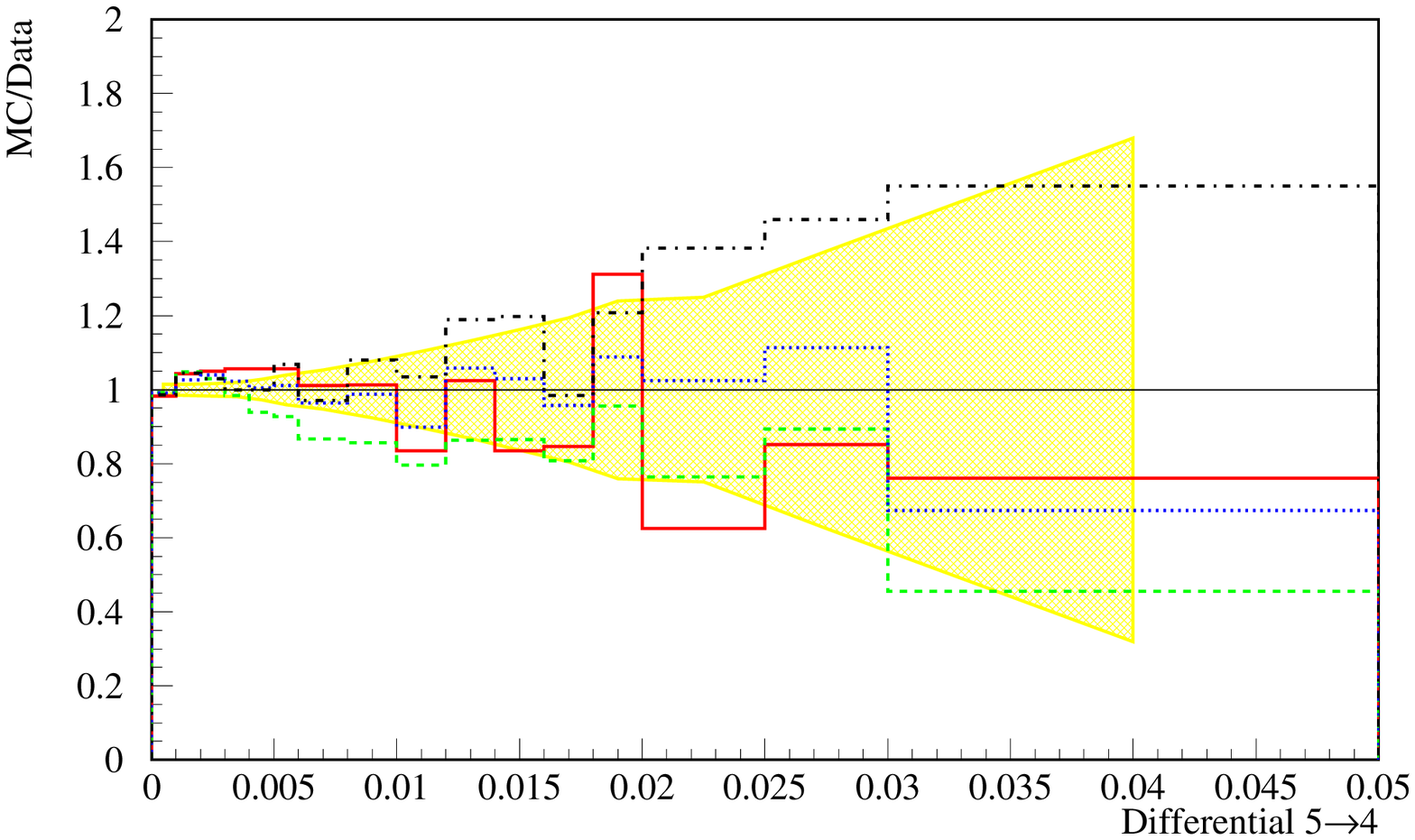}
\end{tabular}
\end{center}
\vspace*{-7mm}
\caption{\label{DJRZ} Differential $3\to 2$, $4\to3$, and
         $5\to 4$ jet rates in the Durham algorithm at the $Z$--pole.
	 DELPHI data (points) are compared to results (curves) of
	 parton shower Monte Carlo generators. The shaded
	 regions denote the size of the experimental errors.}
\end{figure}

In Fig.~\ref{DJRZ} we depict the differential jet rates at the $Z$--pole as
functions of the variable $y_n$, which is the value of $\ycut$ at which an
$n$-jet event becomes an $(n-1)$-jet event.
Clearly, all three event generators depicted here reproduce the shape
of the distributions: deviations are on the level of at most $20\%$ in
the statistically significant bins. In general, {\tt APACIC++} tends to
underestimate the first bins of the $3\to 2$ and $4\to 3$ distributions with
an overshoot in the higher bins. This behaviour is somewhat reversed for the
$5\to 4$ distribution.

\begin{figure}
\begin{center}
\begin{tabular}{cc}
\includegraphics[width=6.5cm,height=9.5cm]{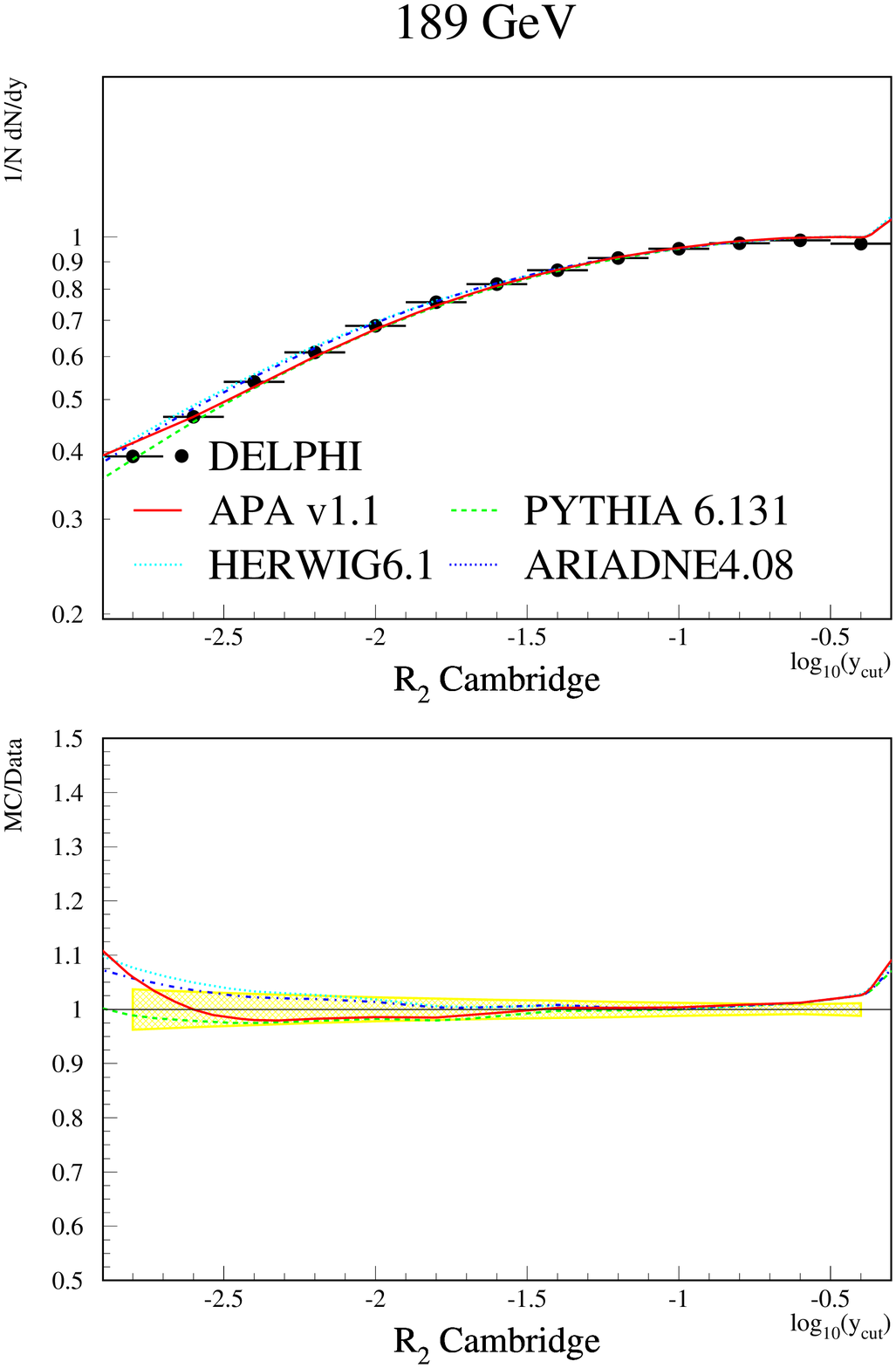} &
\includegraphics[width=6.5cm,height=9.5cm]{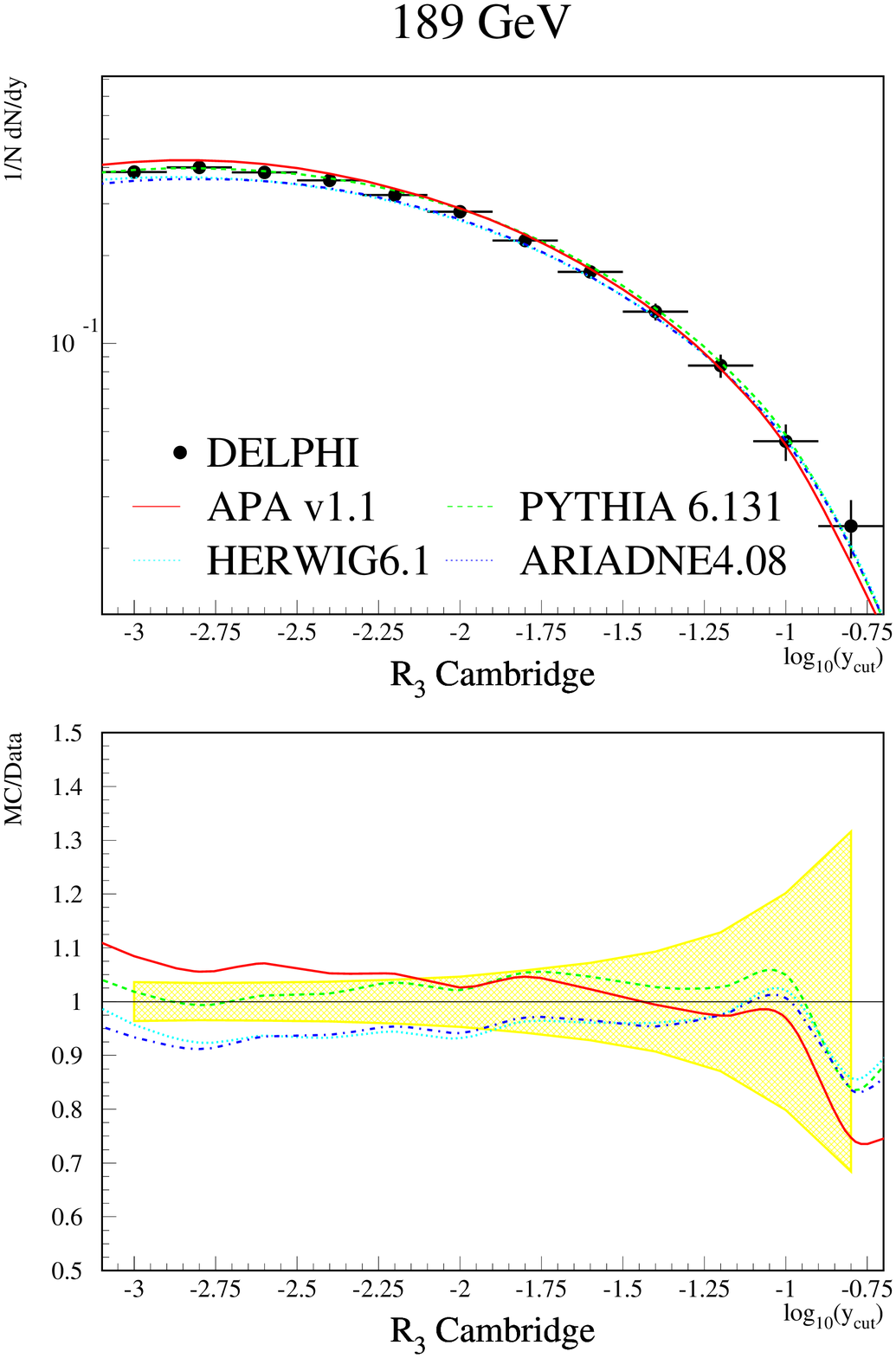}
\end{tabular}
\end{center}
\vspace*{-7mm}
\caption{\label{IJR189} Integrated 2-- and 3--jet rates defined by
         the Cambridge algorithm at ${\sqrt s}=189$~GeV. Note that
         the jet rates predicted by {\tt APACIC++} are in good
         agreement with the experimental ones in the regime of the
         matrix elements, i.e.\ to the right of $\log_{10}\yini=-2.4$}
\end{figure}

Integrated jet rates taken at a c.m.\ energy of 189 GeV,
defined here by the Cambridge algorithm \cite{Dokshitzer:1997in},
are displayed in
Fig.~\ref{IJR189}. They demonstrate that {\tt APACIC++} extrapolates
correctly to higher energies with all parameters fixed at the $Z$--pole.

\begin{figure}
\begin{center}
\begin{tabular}{cc}
\includegraphics[width=6.5cm,height=6.5cm]{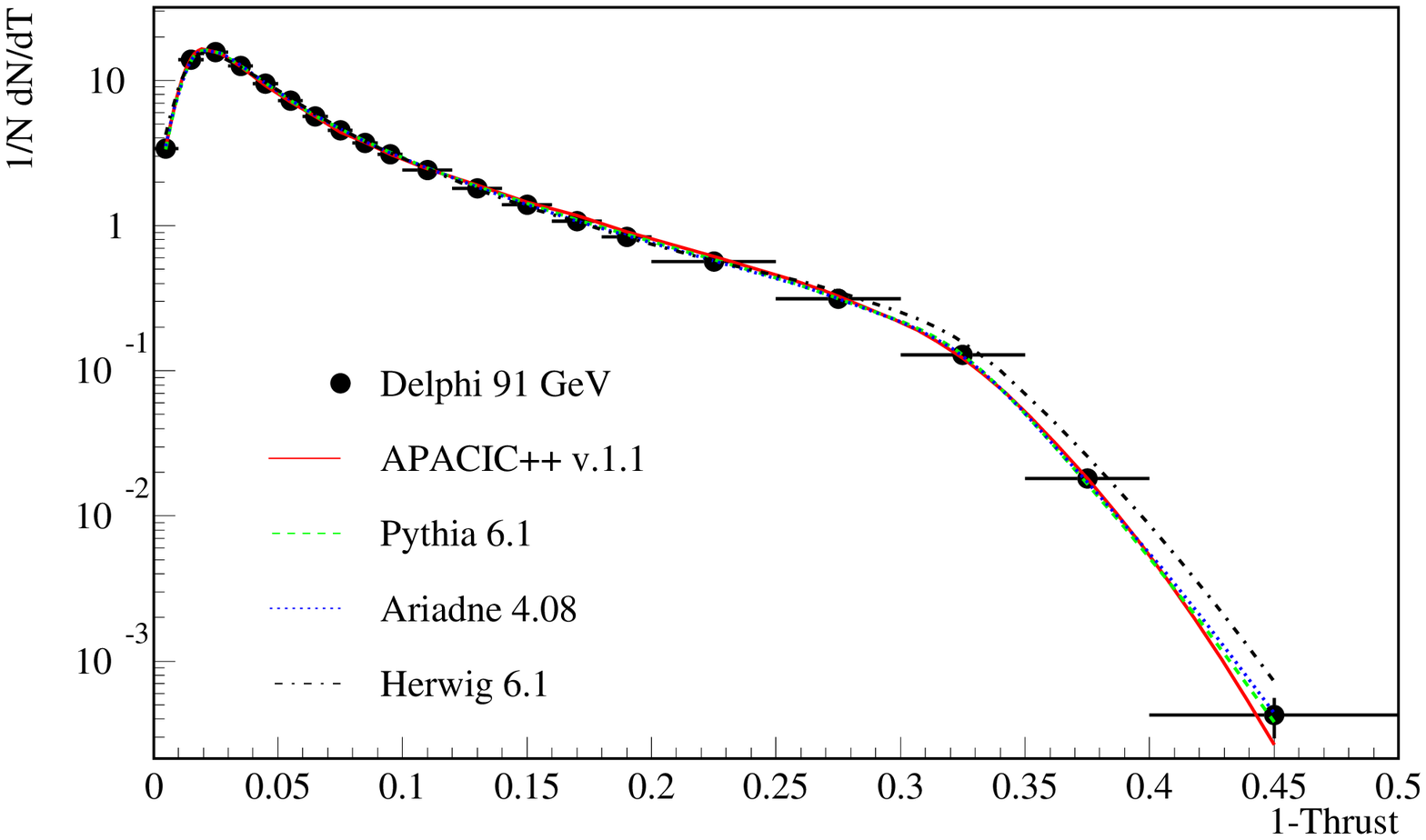} &
\includegraphics[width=6.5cm,height=6.5cm]{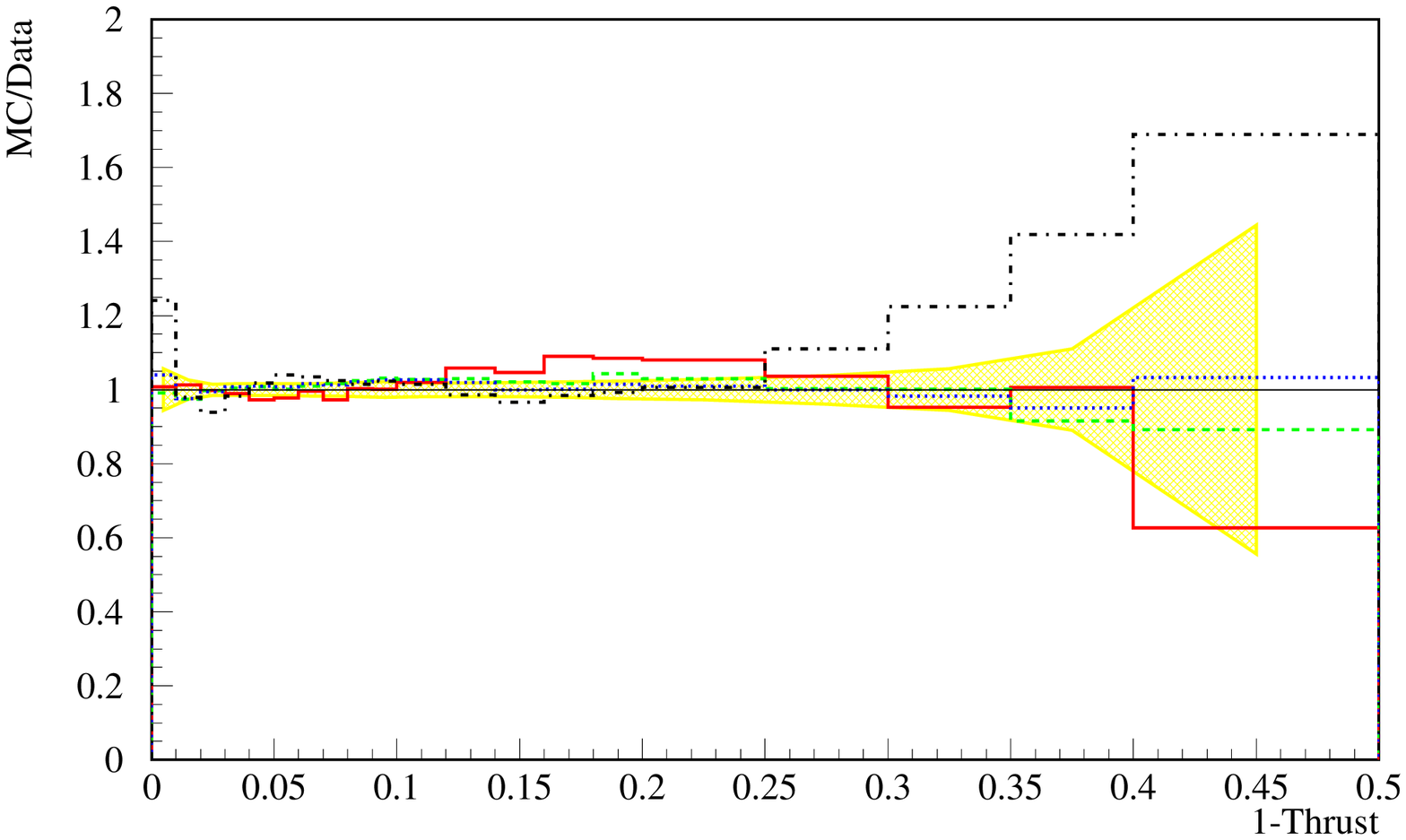} \\
\includegraphics[width=6.5cm,height=6.5cm]{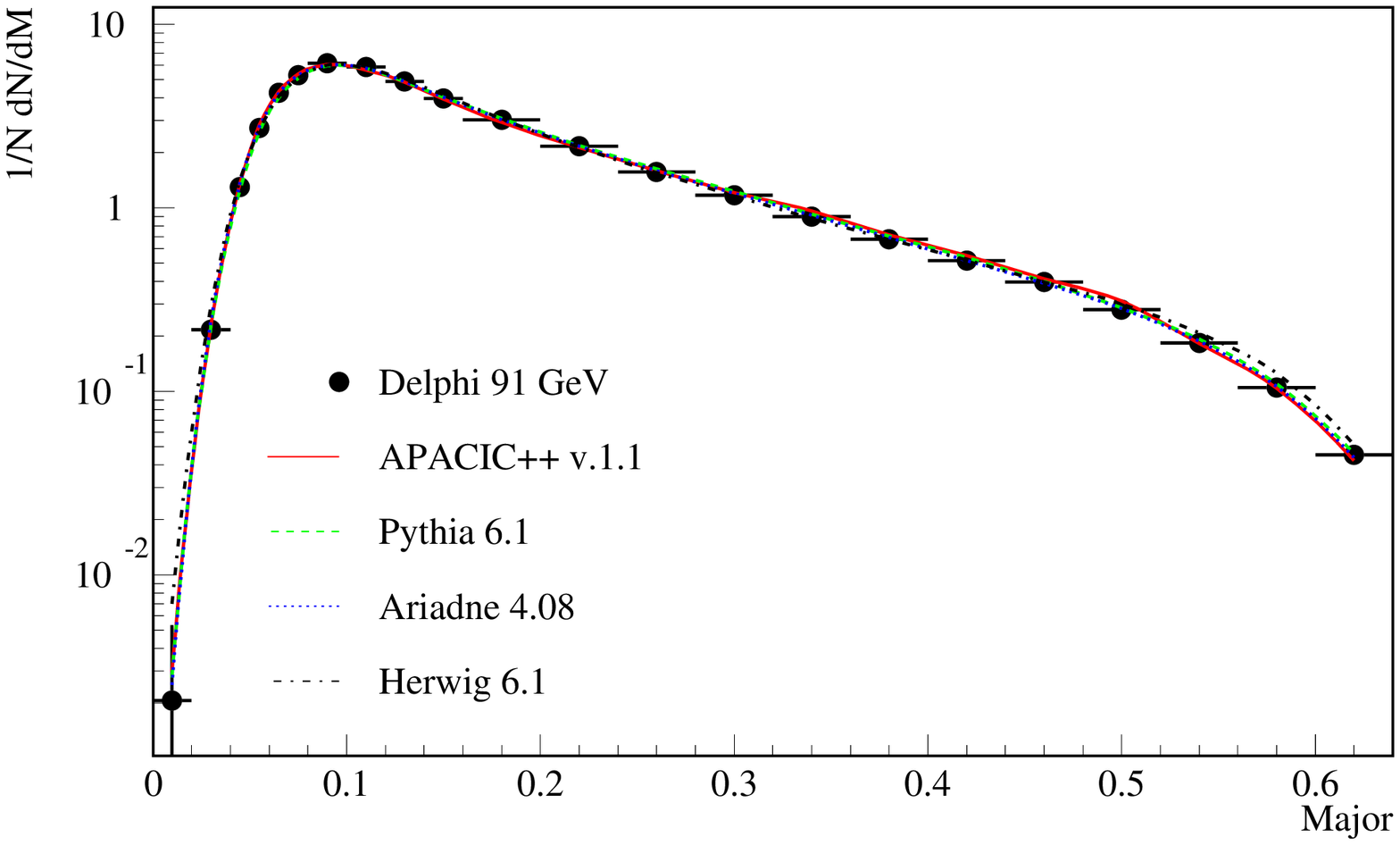}  &
\includegraphics[width=6.5cm,height=6.5cm]{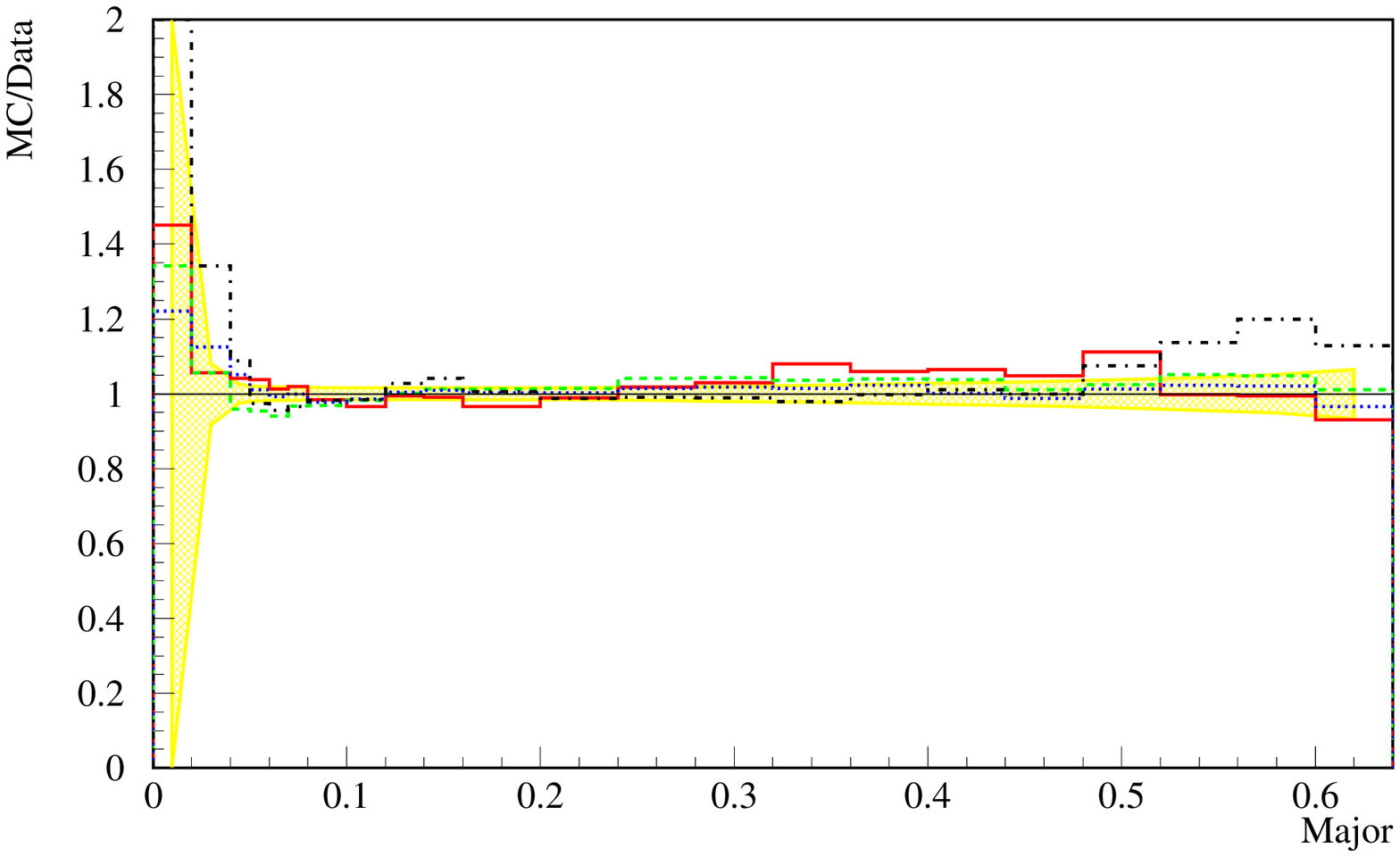}  \\
\includegraphics[width=6.5cm,height=6.5cm]{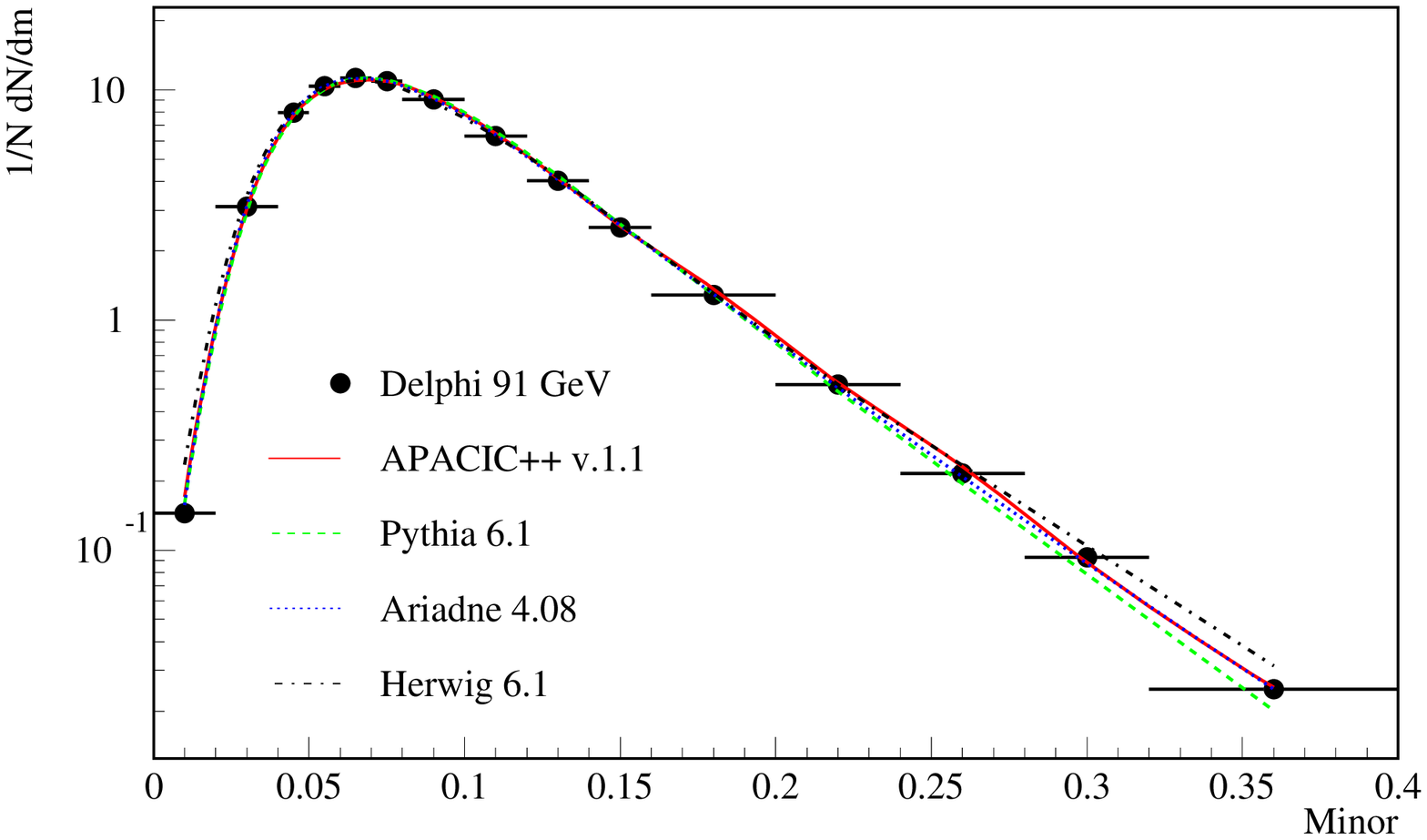}  &
\includegraphics[width=6.5cm,height=6.5cm]{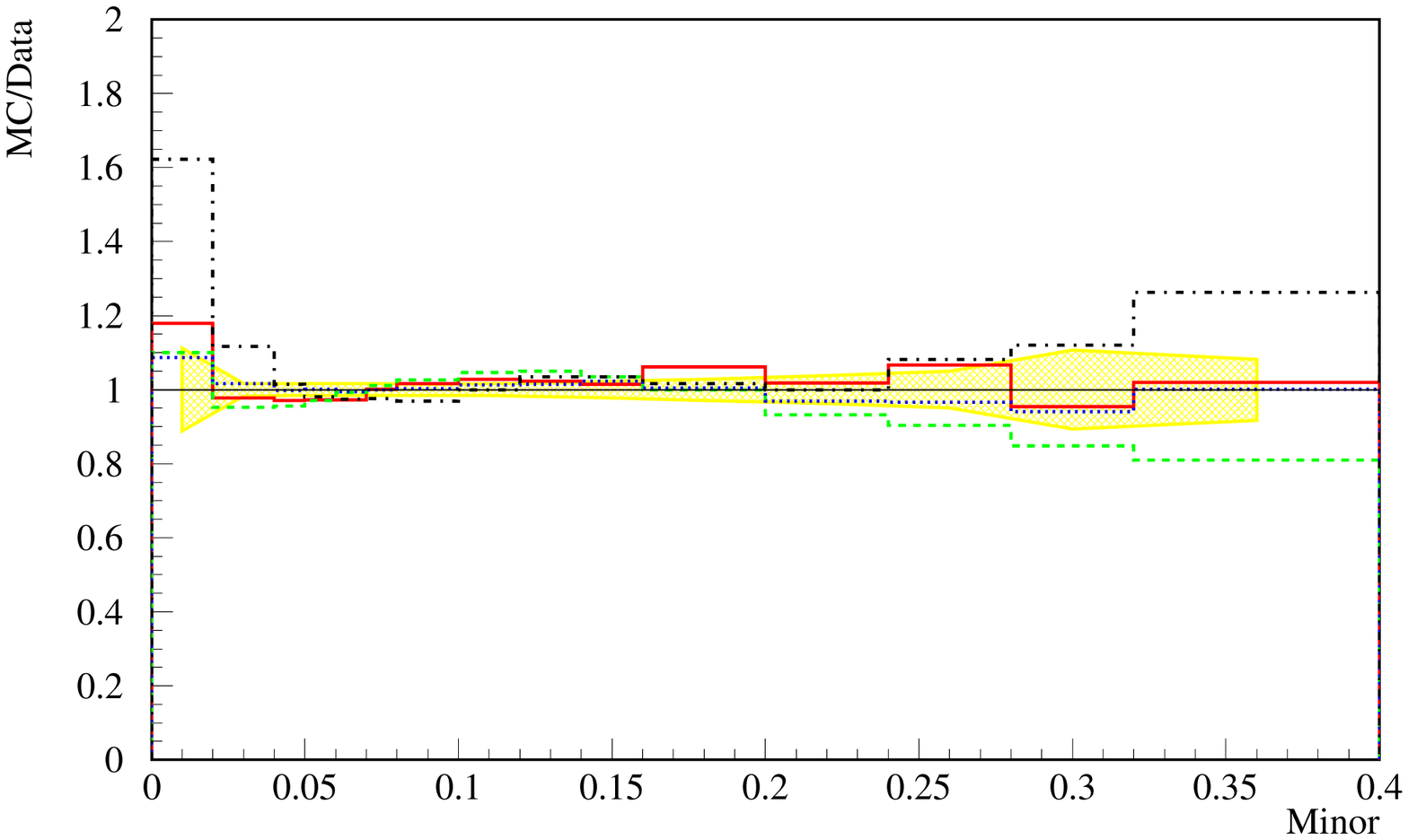}
\end{tabular}
\end{center}
\vspace*{-7mm}
\caption{\label{EvSh} Some event shape (thrust, major and minor)
         distributions at the $Z$--pole.}
\end{figure}

To show that the approach outlined above does indeed
reproduce not only the correct number of jets but also the overall
shape of the events, we display some event shapes taken at the
$Z$--pole (Fig.~\ref{EvSh}).

\begin{figure}
\begin{center}
\begin{tabular}{ccc}
\includegraphics[width=6.5cm,height=6.5cm]{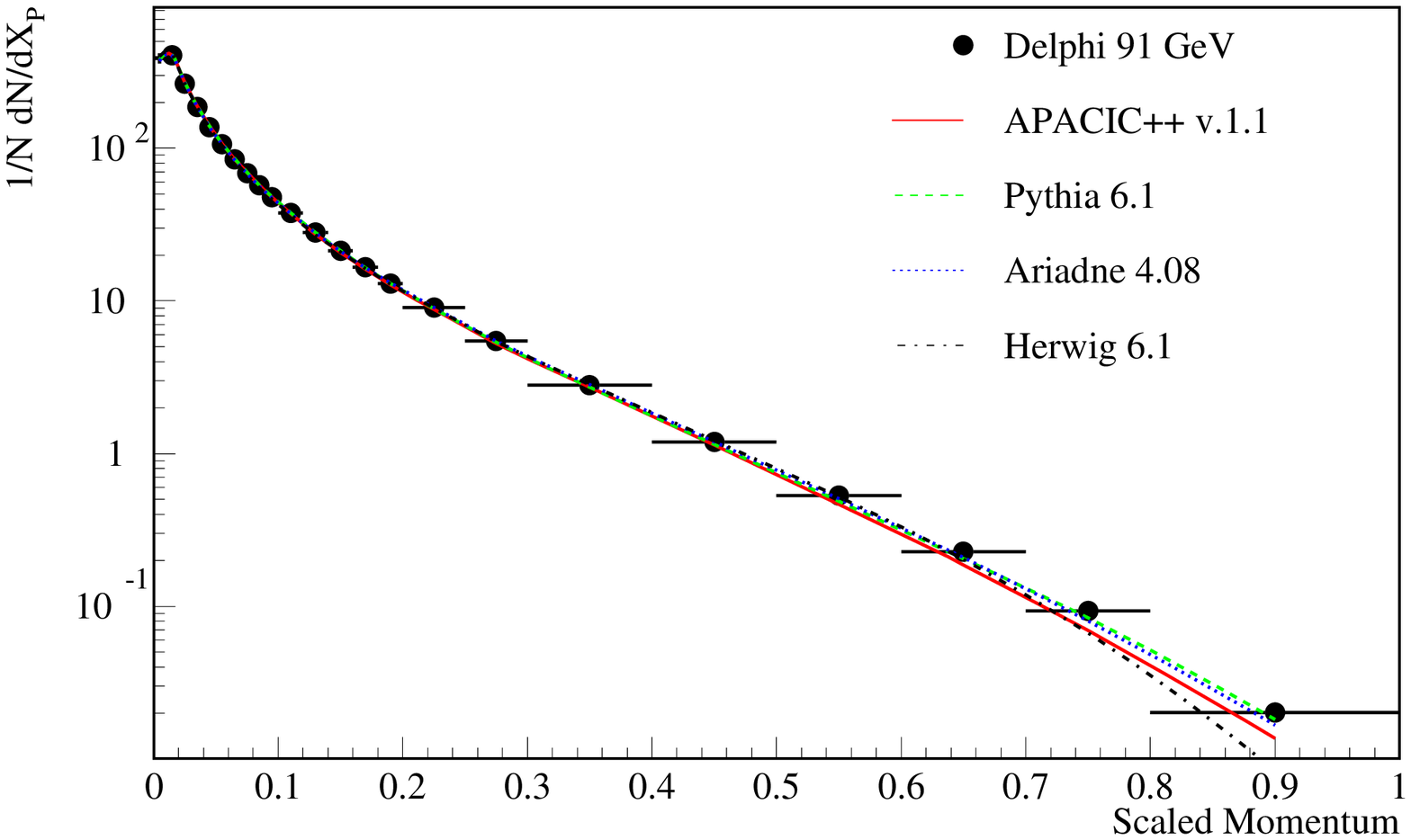} &
\includegraphics[width=6.5cm,height=6.5cm]{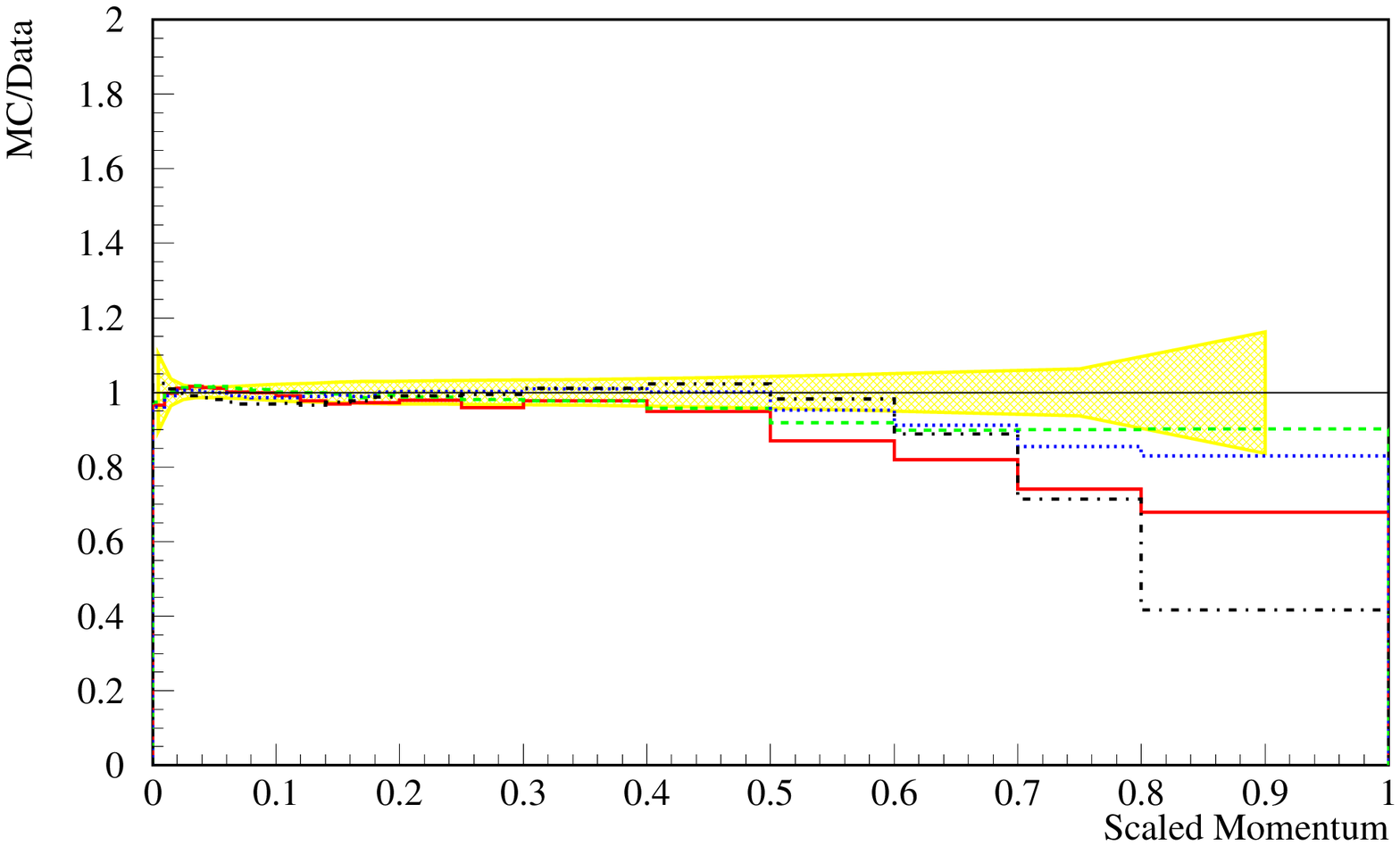} \\
\includegraphics[width=6.5cm,height=6.5cm]{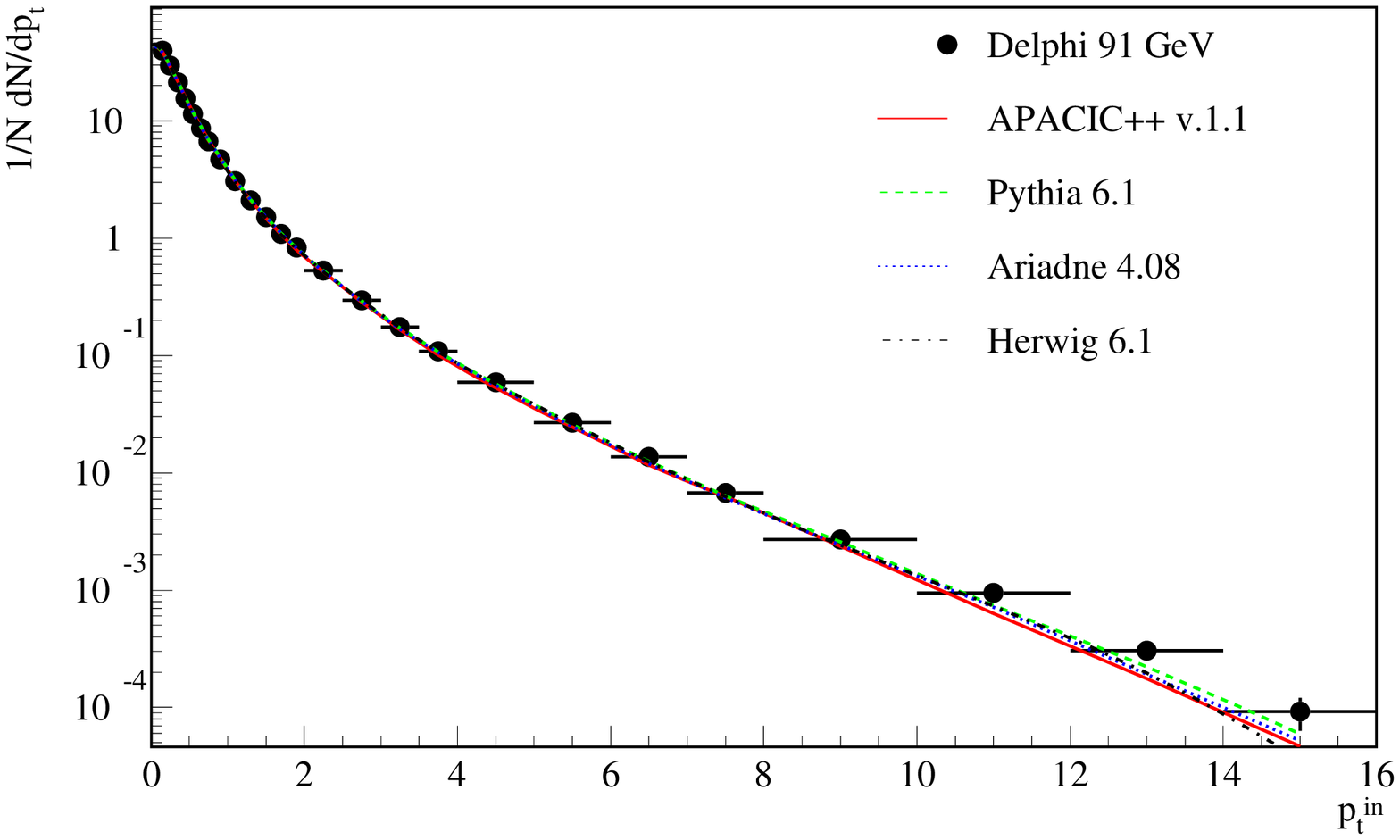}          &
\includegraphics[width=6.5cm,height=6.5cm]{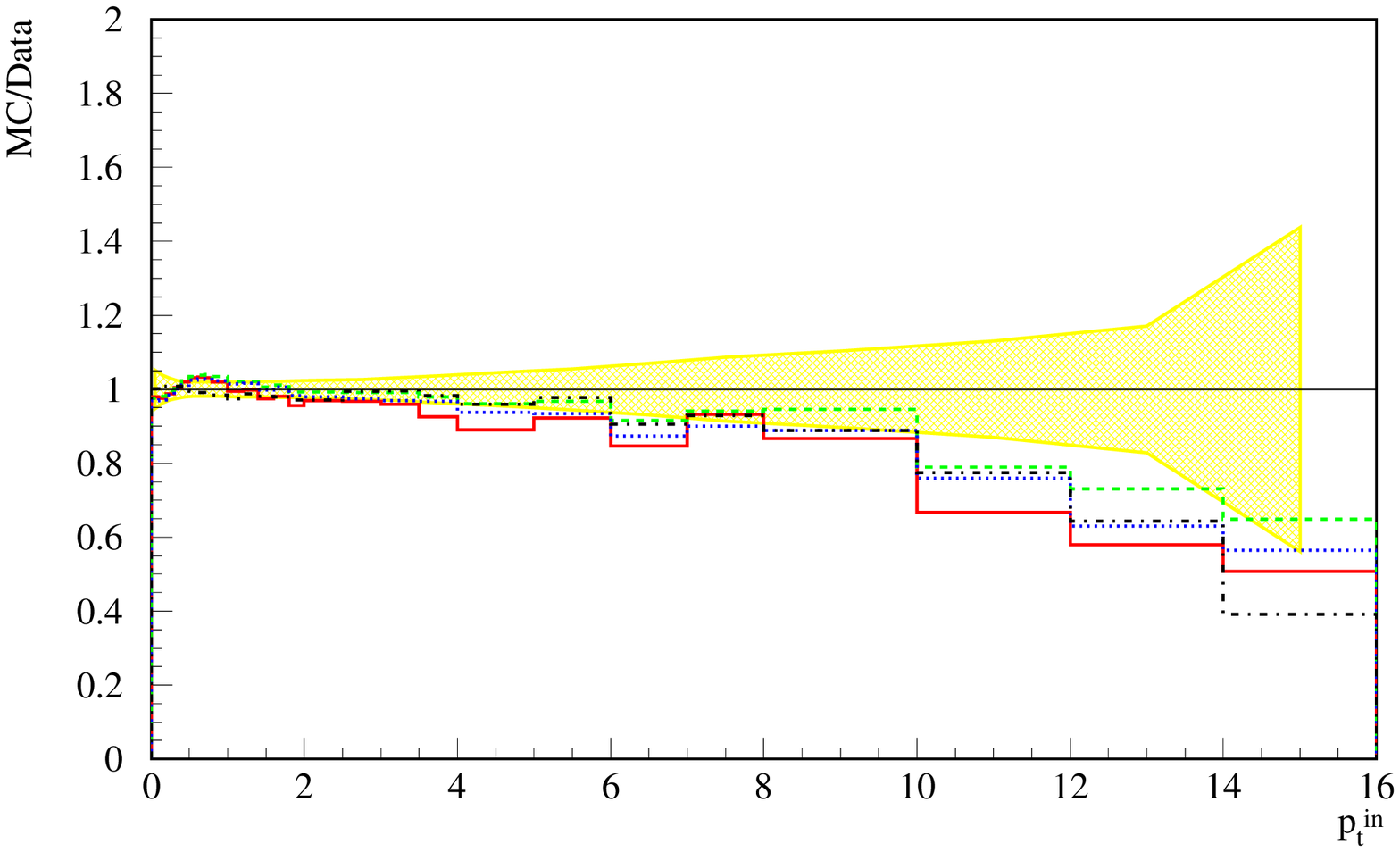}          \\
\includegraphics[width=6.5cm,height=6.5cm]{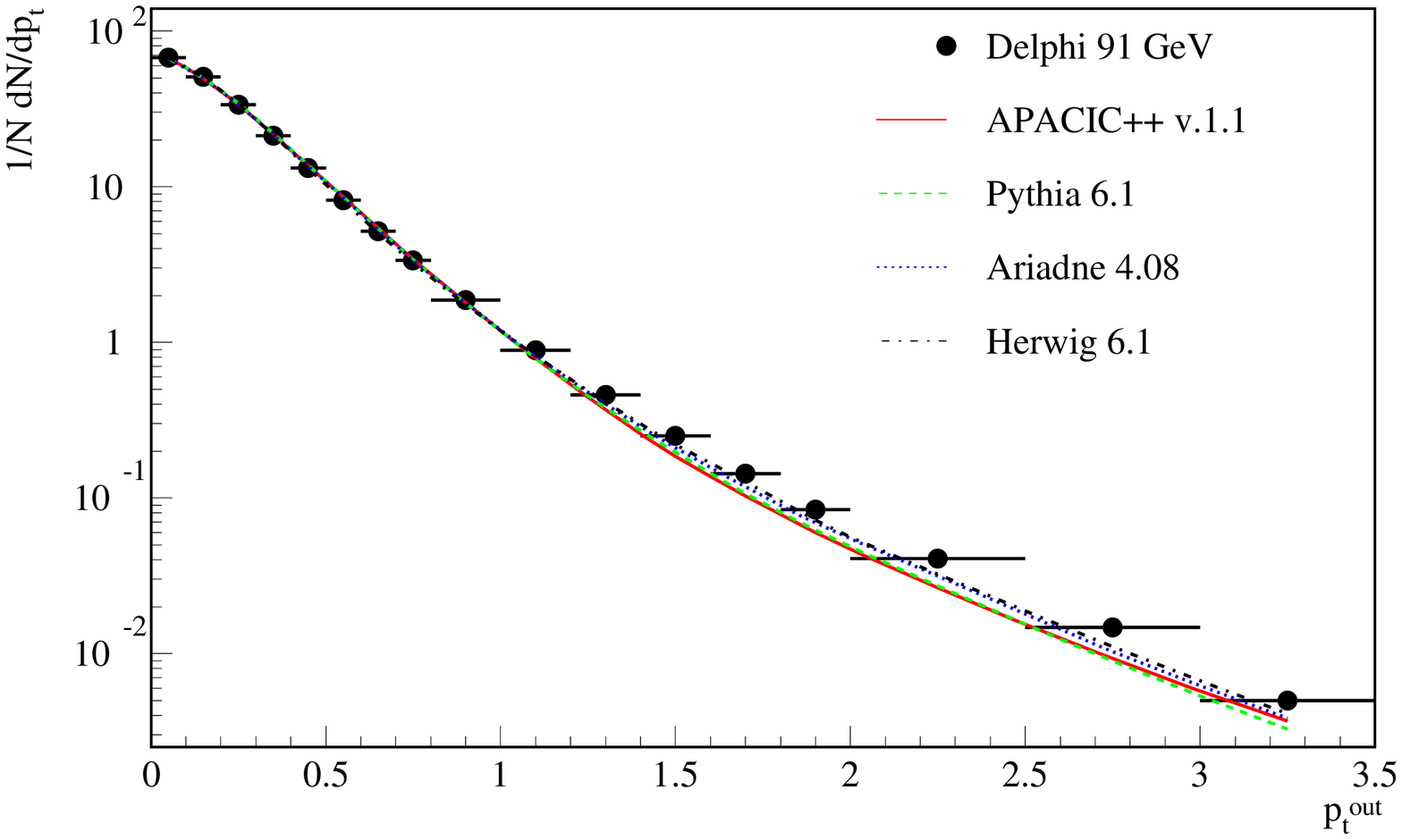}         &
\includegraphics[width=6.5cm,height=6.5cm]{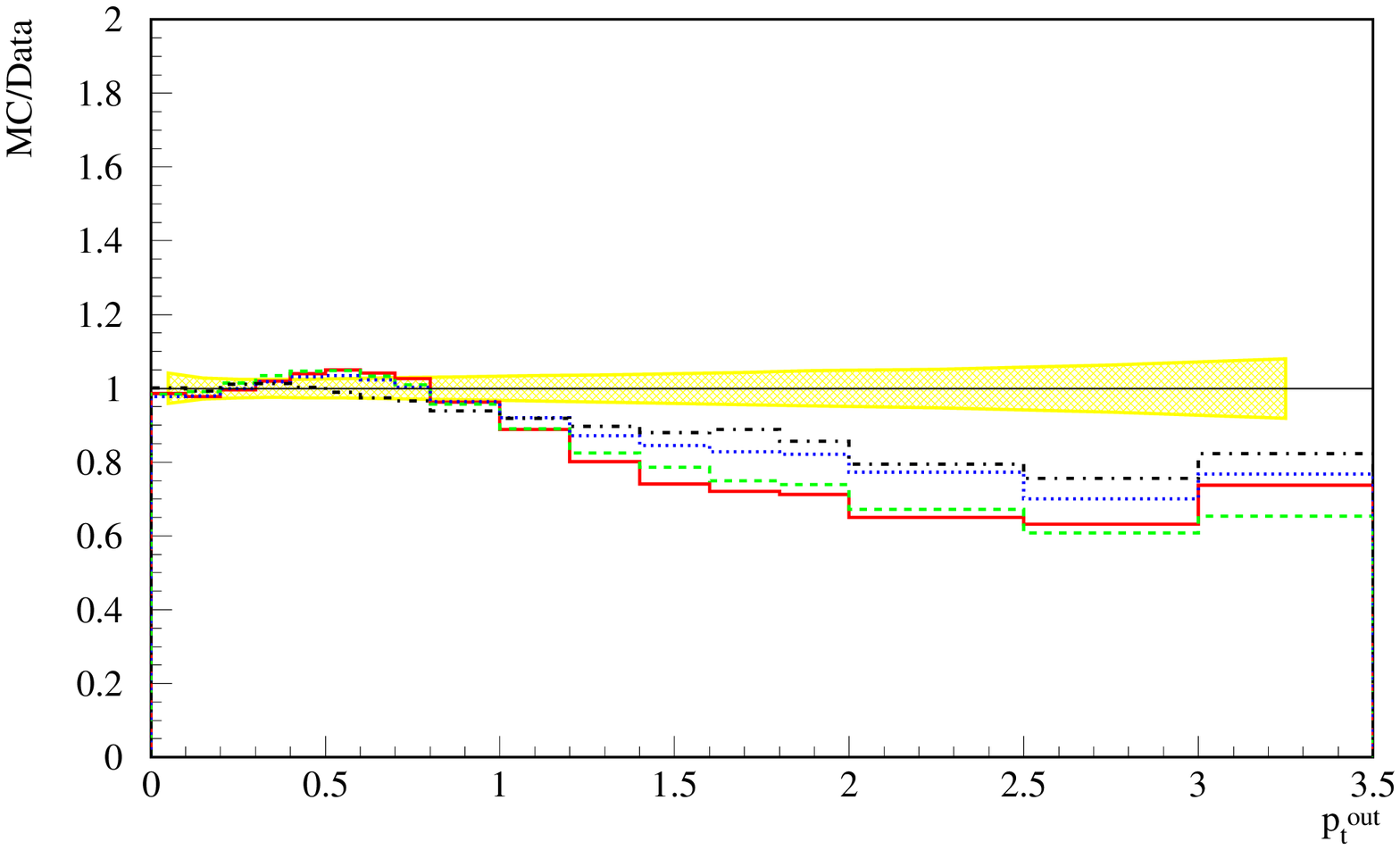}
\end{tabular}
\end{center}
\vspace*{-7mm}
\caption{\label{Moms} Scaled-momentum $(x=2p/{\sqrt s})$,
         $p_t^{\rm in}$ and $p_t^{\rm out}$
         spectra at the $Z$--pole.}
\end{figure}

In Fig.~\ref{Moms} we depict some momentum spectra. Here, all the event
generators tend to underestimate the high-momentum regions. Given the
fact that the overall shapes of the events tend to be reproduced
fairly well by the generators, one is tempted to conclude that
this reflects a lack of particle multiplicity in the high-momentum regions.

However, it should be stressed that the error bands in the right-hand
plots consists of experimental errors -- statistical and systematical
-- only. Monte--Carlo errors of the event generators are not included.
To give some idea of the relative size of these errors, the
numbers of events for the plots at 91 GeV are listed in Table~\ref{NoEv}.
\begin{table}
\begin{center}
\begin{tabular}{|c|c|c|c|c|}
\hline
DELPHI & Ariadne & Herwig & Pythia  & APACIC++ \\\hline
350000 & 2000000 & 250000 & 2000000 & 500000   \\\hline
\end{tabular}
\caption{\label{NoEv} Number of events used to generate the plots.}
\end{center}
\end{table}

\section{Comments/Conclusions}\label{sec_conc}
\begin{itemize}
\item Modified matrix elements plus vetoed parton showers, interfaced
at some value $\yini$ of the $k_T$-resolution parameter, provide
a convenient way to describe simultaneously the hard multi-jet and
jet fragmentation regions.
\item The matrix element modifications are coupling-constant
and Sudakov weights computed directly from the $k_T$-clustering sequence,
which also serves to define the initial conditions for the parton showers.
\item Dependence on $\yini$ is cancelled to NLL accuracy by
vetoing $y_{ij}>\yini$ in the parton showers.
\item  This prescription avoids double-counting problems
and missed phase-space regions.
\item In principle one needs the tree--level matrix elements
$|{\cal M}_{n,i}|^2$ for $\ycut>\yini$ at all values of the
parton multiplicity $n$.
In practice, if we have $n\le N$, then $\yini$ must be chosen
large enough for $R_{n>N}(\yini)$ to be negligible.
\item An approximate version of this approach (with $N=5$)
has been implemented
in the event generator {\tt APACIC++}~\cite{Krauss:1999fc}.
The results look promising: a rather good description of multi-jet
observables can be achieved, and residual dependence on $\yini$ is weak.
\item It should be possible to extend this approach to lepton-hadron and
hadron-hadron collisions. In particular, the procedure discussed
in Sects.~\ref{sec_modme} and \ref{sec_vetps} can be extended to
deep-inelastic lepton-hadron scattering by using the corresponding
calculation of multi-jet rates performed in Ref.~\cite{Catani:1992zp}.
\item Extension to NLO along the lines of
Refs.~\cite{Friberg:1999fh,Collins:2000qd,Potter:2001an,Dobbs:2001gb}
may also be possible.
\end{itemize}

Taken together, the results show sufficient agreement with data to conclude
that this approach to combining matrix elements and parton showers is
successful and merges the benefits of both in a rather simple way.
This approach can also be used to introduce corrections due to the finite mass
of light (with respect to the c.m. energy) quarks, by combining the
massive-quark matrix elements with the corresponding angular-ordered parton
shower~\cite{Marchesini:1990yk}.
It has to be mentioned, however, that some significant
deviations from the data remain.
Therefore,  additional improvements such as the inclusion of NLO matrix
elements seem to be necessary to achieve better agreement.

\section*{Acknowledgments}
The authors are grateful to U.\ Flagmeyer, M.J.\ Costa Mezquita
and the DELPHI collaboration for providing the plots and for helpful
discussions.

BRW thanks CERN, and FK and RK thank the Technion, Israel Institute of
Technology, for kind hospitality while part of this work was done.
FK acknowledges DAAD for funding.

This work was supported in part by the UK Particle Physics and
Astronomy Research Council and by the EU Fourth Framework Programme
`Training and Mobility of Researchers', Network `Quantum Chromodynamics
and the Deep Structure of Elementary Particles',
contract FMRX-CT98-0194 (DG 12 - MIHT).

\end{document}